\documentclass[twocolumn,pre,aps,showpacs,showkeys,amsmath]{revtex4}
\usepackage{graphicx}
\usepackage{adjustbox}
\usepackage{multirow}

\begin{document}

\title{Disynaptic Effect of Hilar Cells on Pattern Separation in A Spiking Neural Network of Hippocampal Dentate Gyrus}
\author{Sang-Yoon Kim}
\email{sykim@icn.re.kr}
\author{Woochang Lim}
\email{wclim@icn.re.kr}
\affiliation{Institute for Computational Neuroscience and Department of Science Education, Daegu National University of Education, Daegu 42411, Korea}

\begin{abstract}
We investigate the disynaptic effect of the hilar cells on pattern separation in a spiking neural network of the hippocampal dentate gyrus (DG). The principal granule cells (GCs) in the DG perform pattern separation, transforming similar input patterns into less-similar output patterns. In our DG network, the hilus consists of excitatory mossy cells (MCs) and inhibitory HIPP (hilar perforant path-associated) cells. Here, we consider the disynaptic effects of the MCs and the HIPP cells on the GCs, mediated by the inhibitory basket cells (BCs) in the granular layer; MC $\rightarrow$ BC $\rightarrow$ GC and HIPP $\rightarrow$ BC $\rightarrow$ GC. Disynaptic inhibition from the MCs tends to decrease the firing activity of the GCs. On the other hand, the HIPP cells disinhibit the intermediate BCs, which leads to increasing the activity of the GCs. By changing the synaptic strength $K^{\rm (BC, X)}$ [from the presynaptic X (= MC or HIPP) to the postsynaptic BC] from the default value ${K^{\rm (BC, X)}}^*$, we study the change in the pattern separation degree ${\cal S}_d$. When decreasing $K^{\rm (BC, MC)}$ or independently increasing $K^{\rm (BC, HIPP)}$ from their default values, ${\cal S}_d$ is found to decrease (i.e., pattern separation is reduced). On the other hand, as $K^{\rm (BC, MC)}$ is increased or independently $K^{\rm (BC, HIPP)}$ is decreased from their default values, pattern separation becomes enhanced (i.e., ${\cal S}_d$ increases). In this way, the disynaptic effects of the MCs and the HIPP cells on the pattern separation are opposite ones. Thus, when simultaneously varying both $K^{\rm (BC, MC)}$ and $K^{\rm (BC, HIPP)}$, as a result of balance between the two competing disynaptic effects of the MCs and the HIPP cells, ${\cal S}_d$ forms a bell-shaped curve with an optimal maximum at their default values. Moreover, we also investigate population and individual behaviors of the sparsely synchronized rhythm of the GCs, and find that the amplitude measure ${\cal M}_a$ (representing population synchronization degree) and the random-phase-locking degree ${\cal L}_d$ (denoting individual activity degree) are strongly correlated with the pattern separation degree ${\cal S}_d$. Consequently, the larger the  synchronization and the random phase-locking degrees of the sparsely synchronized rhythm is, the more the pattern separation becomes enhanced.
\end{abstract}


\maketitle

\section{Introduction}
\label{sec:INT}
The hippocampus, consisting of the dentate gyrus (DG) and the areas CA3 and CA1, plays important roles in memory formation, storage, and retrieval
(e.g., episodic and spatial memory) \cite{Gluck,Squire}. Particularly, the area CA3 has been considered as an autoassociative network, due to extensive recurrent collateral synapses between the pyramidal cells in the CA3 \cite{Marr,Will,Mc,Rolls1,Rolls2a,Rolls2b,Treves1,Treves2,Treves3,Oreilly}. This autoassociative network operates in the two storage and recall modes. In the storage mode, it stores input patterns in plastic collateral synapses between the pyramidal cells. In the recall mode, when an incomplete partial version of ``cue'' pattern is presented, activity of pyramidal cells propagates through the previously-strengthened pathways and reinstates the complete stored pattern, which is called the pattern completion.

Storage capacity of the autoassociative network implies the number of distinct patterns that can be stored and accurately retrieved.
Such storage capacity may be increased if the input patterns are sparse (containing few active elements in each pattern) and (non-overlapping) orthogonalized
(active elements in one pattern are unlikely to be active in other patterns). This process of transforming a set of input patterns into sparser and orthogonalized patterns is called pattern separation \cite{Marr,Will,Mc,Rolls1,Rolls2a,Rolls2b,Treves1,Treves2,Treves3,Oreilly,Schmidt,Rolls3,Knier,Myers1,Myers2,Myers3,Scharfman,Yim,Chavlis,Kassab,PS1,PS2,PS3,PS4,PS5,PS6,PS7}.

The DG is the gateway to the hippocampus, and its excitatory granule cells (GCs) receive excitatory inputs from the entorhinal cortex (EC)
via the perforant paths (PPs). As a pre-processor for the CA3, the principal GCs perform pattern separation on the input patterns from the EC by sparsifying and orthogonalizing them, and send the pattern-separated outputs to the pyramidal cells in the CA3 through the mossy fibers (MFs)
\cite{Treves3,Oreilly,Schmidt,Rolls3,Knier,Myers1,Myers2,Myers3,Scharfman,Yim,Chavlis,Kassab}.
Then, the sparse, but strong MFs play a role of ``teaching inputs,'' causing synaptic plasticity between the pyramidal cells in the CA3.
Thus, a new pattern may be stored in modified synapses. In this way, pattern separation in the DG may facilitate pattern storage in the CA3.

The whole GCs are grouped into the lamellar clusters \cite{Cluster1,Cluster2,Cluster3,Cluster4}. In each lamella, there exists one inhibitory basket cell (BC)
together with excitatory GCs. In the process of pattern separation, the GCs exhibit sparse firing activity through the winner-take-all competition \cite{WTA1,WTA2,WTA3,WTA4,WTA5,WTA6,WTA7,WTA8,WTA9,WTA10}. Only strongly active GCs survive under the feedback inhibition of the BC.
The sparsity (resulting from the winner-take-all competition) has been considered to improve the pattern separation
\cite{Treves3,Oreilly,Schmidt,Rolls3,Knier,Myers1,Myers2,Myers3,Scharfman,Chavlis,Kassab}.

In this paper, we consider a spiking neural network of the hippocampal DG, and investigate the disynaptic effect of the hilar cells on pattern separation.
Our work is in contrast to the previous work on the monosynaptic effect of the hilar cells on the pattern separation \cite{Myers1}.
In our DG network, the hilus is composed of two kinds of hilar cells: excitatory mossy cells (MCs) and inhibitory HIPP (hilar perforant path-associated) cells.
We are focused on the disynaptic effect of the MCs and the HIPP cells on the GCs (performing pattern separation), mediated by the inhibitory BCs;
MC $\rightarrow$ BC $\rightarrow$ GC and HIPP $\rightarrow$ BC $\rightarrow$ GC, in contrast to their monosynaptic effect on pattern separation
(MC $\rightarrow$ GC and HIPP $\rightarrow$ GC) \cite{Myers1}. In our case, disynaptic inhibition from the MCs tends to reduce the firing activity of the GCs, while the HIPP cells have tendency of increasing the firing activity of the GCs by disinhibiting the BCs; these disynaptic effects are opposite to the monosynaptic effects.

By varying the synaptic strength $K^{\rm (BC, X)}$ [from the presynaptic X (= MC or HIPP) to the postsynaptic BC] from the default value ${K^{\rm (BC, X)}}^*$,
we study the change in the activation degree $D_a^{\rm (GC)}$ of the GCs and the pattern separation degree ${\cal S}_d$.
As $K^{\rm (BC, MC)}$ is decreased, or independently $K^{\rm (BC, HIPP)}$ is increased from their default values, pattern separation is reduced (i.e.,
${\cal S}_d$ decreases). In contrast, when increasing $K^{\rm (BC, MC)}$ or independently decreasing $K^{\rm (BC, HIPP)}$ from their default values,
pattern separation becomes enhanced (i.e., ${\cal S}_d$ increases). In this way, the disynaptic effect of the MCs are opposite to that of the HIPP cells.

As a result of balance between the two competing disynaptic effects of the MCs and the HIPP cells, when simultaneously changing both $K^{\rm (BC, MC)}$ and $K^{\rm (BC, HIPP)}$, ${\cal S}_d$ is found to form a bell-shaped curve with an optimal maximum at their default values (i.e., the pattern separation becomes the best at the default values of our network).  In contrast, the activation degree $D_a^{\rm (GC)}$ of the GCs forms a well-shaped curve with an optimal minimum at their default values. Hence, in the sparsest case for the activity of the GCs, the pattern separation degree ${\cal S}_d$  becomes maximal.

During the pattern separation, sparsely synchronized rhythm appears in the population of the GCs.
In the combined case of simultaneously changing $K^{\rm (BC, MC)}$ and $K^{\rm (BC, HIPP)}$, we also investigate the population and the individual activities of the sparsely synchronized rhythm of the GCs. The amplitude measure ${\cal M}_a$ (denoting population synchronization degree) and the random-phase-locking degree ${\cal L}_d$ (representing individual activity degree) are found to be strongly correlated with the pattern separation degree ${\cal S}_d$. Consequently, the larger the  synchronization and the random phase-locking degrees of the sparsely synchronized rhythm is, the better the pattern separation becomes.

Finally, for comparison, we study the monosynaptic effect of the MCs and the HIPP cells on pattern separation.
In contrast to the disynaptic case, the MCs and the HIPP cells provide direct excitation and inhibition to the GCs, respectively.
In the combined case of simultaneously varying both $K^{\rm (GC, MC)}$ and $K^{\rm (GC, HIPP)}$,
${\cal S}_d$ ($D_a^{\rm (GC)}$) is found to form a well-shaped (bell-shaped) curve with optimal minimum (maximum), in contrast to disynaptic case with the up-down flipped curves. In this way, the monosynaptic effect is opposite to the disynaptic effect.

This paper is organized as follows. In Sec.~\ref{sec:DGN}, we describe a spiking neural network for pattern separation in the hippocampal DG. Then, in the main Sec.~\ref{sec:PS}, we investigate the disynaptic effects of the MCs and the HIPP cells on pattern separation and its association with the population and the individual activities of the sparsely synchronized rhythm of the GCs. Finally, we give summary and discussion in Sec.~\ref{sec:SUM}.

\section{Spiking Neural Network for The Pattern Separation in The Dentate Gyrus}
\label{sec:DGN}
In this section, we describe our spiking neural network for the pattern separation in the DG.
Based on the anatomical and the physiological properties given in \cite{Myers1,Chavlis}, we developed the DG spiking neural network in the work for the winner-take-all competition \cite{WTA} and the sparsely synchronized rhythms \cite{SSR}.
In the present work, we start with the DG network for the sparsely synchronized rhythm \cite{SSR}, and modify it for the study on the disynaptic effect of the hilar cells on pattern separation. There is no disynaptic connections from the HIPP cells to the GCs, mediated by the BCs (HIPP  $\rightarrow$ BC $\rightarrow$ GC) in \cite{SSR}. For our present study, we make such disynaptic connections from the HIPP cells in the DG network for the pattern separation, in addition to the
(pre-existing) disynaptic connections from the MCs, mediated by the BCs (MC  $\rightarrow$ BC $\rightarrow$ GC). To keep the same activation degree $D_a$ (= 5.2 $\%$) of the GCs as in the case of sparsely synchronized rhythm, we make a little changes in the following synaptic strengths: increase in the synaptic strength for HIPP $\rightarrow$ GC and increase in the activity of the GC-MC-BC loop via increasing the synaptic strengths for GC $\rightarrow$ MC and MC $\rightarrow$ BC.
We also determine the synaptic strength for the new disynaptic connection (HIPP $\rightarrow$ BC), based on the information in \cite{BN1,BN2}.
Obviously, our spiking neural network will not capture all the detailed anatomical and physiological complexity of the DG. But, with a limited number of essential elements and synaptic connections in our DG network, disynaptic effect on the pattern separation could be successfully studied. Hence, our spiking neural network model would build a foundation upon which additional complexity may be added and guide further research.

\begin{figure}
\includegraphics[width=1.0\columnwidth]{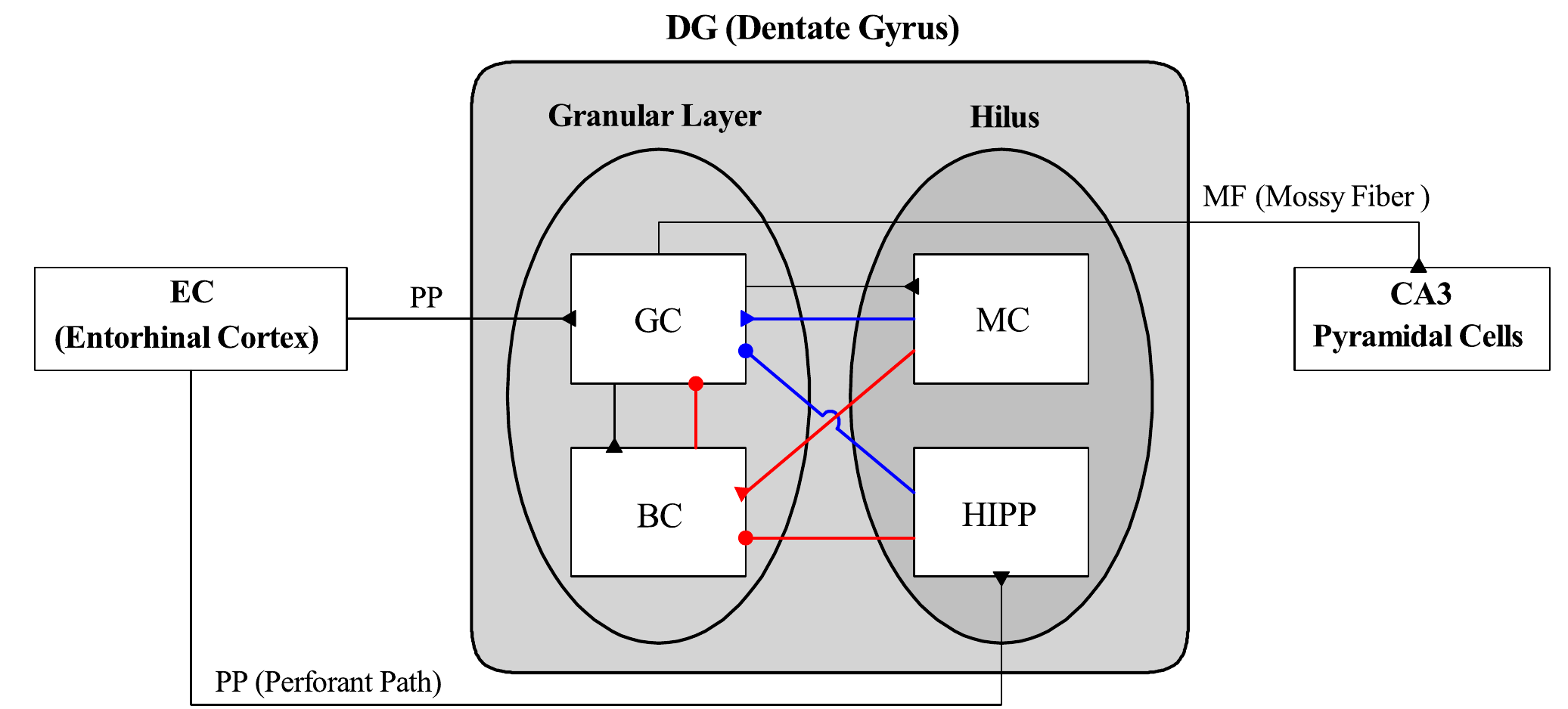}
\caption{Box diagram for the hippocampal dentate gyrus (DG) network. Lines with triangles and circles denote excitatory and inhibitory synapses, respectively. In the DG, there are the granular layer [consisting of GC (granule cell) and BC (basket cell)] and the hilus [composed of MC (mossy cell) and HIPP (hilar perforant path-associated) cell]. The DG receives excitatory input from the EC (entorhinal cortex) via PPs (perforant paths) and provides its output to the CA3 via MFs (mossy fibers). Red and blue lines represent disynaptic and monosynaptic connections into GCs, respectively.
}
\label{fig:DGN}
\end{figure}

\subsection{Architecture of The Spiking Neural Network of The Dentate Gyrus}
\label{subsec:SNN}
Figure \ref{fig:DGN} shows the box diagram for the DG network for our study on patten separation. In the DG, there exist the granular layer (consisting of the excitatory GCs and the inhibitory BCs) and the hilus (composed of the excitatory MCs and the inhibitory HIPP cells). This DG receives the input from the external EC via the PPs and projects its output to the CA3 via the MFs. Inside the DG, disynaptic paths from the MCs and the HIPP cells to the GCs, mediated by the BCs, are represented in red color, while direct monosynaptic paths from the  MCs and the HIPP cells are denoted in blue color.

Based on the anatomical information given in \cite{Myers1,Chavlis}, we choose the numbers of the GCs, BCs, MCs, and HIPP cells in the DG and the EC cells and the connection probabilities between them. As in the work for the sparsely synchronized rhythm \cite{SSR}, we develop a scaled-down spiking neural network where the total number of excitatory GCs ($N_{\rm GC}$) is 2,000, corresponding to $\frac {1}{500}$ of the $10^6$ GCs found in rats \cite{ANA1}. The GCs are grouped into the $N_c~(=20)$ lamellar clusters \cite{Cluster1,Cluster2,Cluster3,Cluster4}. Then, in each GC cluster, there are $n_{\rm {GC}}^{(c)}~(=100)$ GCs and one inhibitory BC.
As a result, the number of the BCs ($N_{\rm BC}$) in the whole DG network become 20, corresponding to 1/100 of $N_{\rm GC}$ \cite{GC-BC1,GC-BC2,GC-BC3,GC-BC4,GC-BC5,BN2}. In this way, in each GC cluster, a dynamical GC-BC loop is formed, and the BC (receiving the excitation from all the GCs) provide the feedback inhibition to all the GCs.

The EC layer II projects the excitatory inputs to the GCs and the HIPP cells via the PPs, as shown in Fig.~1 in Ref.~\cite{Myers1}.
The HIPP cells have dendrites extending into the outer molecular layer, where they are targeted by the PPs, together with axons projecting to the outer molecular layer \cite{Myers1,FF-A1,FF-A2,FF-A3}. In this way, the EC cells and the HIPP cells become the excitatory and the inhibitory input sources to the GCs, respectively.
The estimated number of the EC layer II cells ($N_{\rm EC}$) is about 200,000 in rats, which corresponds to 20 EC cells per 100 GCs  \cite{ANA3}.
Hence, we choose $N_{\rm EC}=400$ in our DG network. Also, the activation degree $D_a$ of the EC cells is chosen as 10$\%$ \cite{ANA4}. Thus, we randomly choose 40 active ones among the 400 EC cells. Each active EC cell is modeled in terms of the Poisson spike train with frequency of 40 Hz \cite{ANA5}.
The random-connection probability $p^{\rm (GC,EC)}$ ($p^{\rm (HIPP,EC)}$) from the pre-synaptic EC cells to a post-synaptic GC (HIPP cell) is 20 $\%$ \cite{Myers1,Chavlis}. Thus, each GC or HIPP cell is randomly connected with the average number of 80 EC cells.

Next, we consider the hilus, composed of the excitatory MCs and the inhibitory HIPP cells \cite{Hilus1,Hilus2,Hilus3,Hilus4,Hilus5,Hilus6,Hilus7}.
In rats, the number of MCs ($N_{\rm MC}$) is known to change from 30,000 to 50,000, which corresponds to 3-5 MCs per 100 GCs \cite{ANA1,ANA2}. In our DG network, we choose $N_{\rm MC}=80$. Also, the estimated number of HIPP cells ($N_{\rm HIPP}$) in rats is about 12,000 \cite{ANA2}, corresponding to about 2 HIPP cells per 100 GCs. Hence, we chose $N_{\rm HIPP}=40$ in our DG network. For simplicity, as in \cite{Myers1,Chavlis}, we do not consider the lamellar cluster organization for the hilar cells.

In our DG network, the whole MCs and the GCs in each GC cluster were mutually connected with the same 20 $\%$ random-connection probabilities $p^{\rm (MC,GC)}$ ($\rm GC \rightarrow MC$) and $p^{\rm (GC,MC)}$ ($\rm MC \rightarrow GC$), independently of the GC clusters \cite{Myers1,Chavlis}. In this way, the GCs and the MCs form a dynamical E-E loop. Also, the BC in each GC cluster is randomly connected with the whole MCs with the connection probability $p^{\rm (BC,MC)}=20~\%$, in contrast to the case of sparsely synchronized rhythm \cite{SSR} where all the MCs provide excitation to the BC in each GC cluster \cite{Chavlis}.
In this way, the MCs control the firing activity in the GC-BC loop by providing excitation to both the randomly-connected GCs and BCs.

We also note that each GC in the GC cluster receive inhibition from the randomly-connected HIPP cells with the connection probability $p^{\rm (GC,HIPP)}=20 ~\%$
\cite{Myers1,Chavlis}. Hence, the firing activity of the GCs may be determined through competition between the excitatory inputs from the EC cells and from the MCs and the inhibitory inputs from the HIPP cells.

With the above information on the numbers of the relevant cells and the connection probabilities between them, we develop a one-dimensional ring network for the
pattern separation in the DG, as in the case of sparsely synchronized rhythm and winner-take-all competition in the DG \cite{WTA,SSR}. Due to the ring structure, our network has advantage for computational efficiency, and its visual representation may also be easily made. For the schematic diagrams of the ring networks for the EC, the granular layer and the hilus, refer to Figs.~1(b1)-1(b3) in \cite{WTA}, respectively.

\subsection{Elements in The DG Spiking Neural Network}
\label{subsec:LIF}
As elements of our DG spiking neural network, we choose leaky integrate-and-fire (LIF) neuron models with additional afterhyperpolarization (AHP) currents which  determines refractory periods, like our prior study of cerebellar network \cite{Kim1,Kim2}. This LIF neuron model is one of the simplest spiking neuron models \cite{LIF}. Due to its simplicity, it may be easily analyzed and simulated.

The governing equations for evolutions of dynamical states of individual cells in the $X$ population are as follows:
\begin{eqnarray}
C_{X} \frac{dv_{i}^{(X)}(t)}{dt} &=& -I_{L,i}^{(X)}(t) - I_{AHP,i}^{(X)}(t) + I_{ext}^{(X)} - I_{syn,i}^{(X)}(t), \nonumber \\
& & \;\;\; i=1, \cdots, N_{X},\label{eq:GE}
\end{eqnarray}
where $N_X$ is the total number of cells in the $X$ population, $X=$ GC and BC in the granular layer and $X=$ MC and HIPP in the hilus.
In Eq.~(\ref{eq:GE}), $C_{X}$ (pF) denotes the membrane capacitance of the cells in the $X$ population, and the state of the $i$th cell in the $X$ population at a time $t$ (msec) is characterized by its membrane potential $v_i^{(X)}(t)$ (mV). We note that the time-evolution of $v_i^{(X)}(t)$ is governed by 4 types of currents (pA) into the $i$th cell in the $X$ population; the leakage current $I_{L,i}^{(X)}(t)$, the AHP current $I_{AHP,i}^{(X)}(t)$, the external constant current $I_{ext}^{(X)}$ (independent of $i$), and the synaptic current $I_{syn,i}^{(X)}(t)$. Here, we consider a subthreshold case of $I_{ext}^{(X)}=0$ for all $X$ \cite{Chavlis}.

The 1st type of leakage current $I_{L,i}^{(X)}(t)$ for the $i$th cell in the $X$ population is given by:
\begin{equation}
I_{L,i}^{(X)}(t) = g_{L}^{(X)} (v_{i}^{(X)}(t) - V_{L}^{(X)}),
\label{eq:Leakage}
\end{equation}
where $g_L^{(X)}$ and $V_L^{(X)}$ are conductance (nS) and reversal potential for the leakage current, respectively.
The $i$th cell fires a spike when its membrane potential $v_i^{(X)}$ reaches a threshold $v_{th}^{(X)}$ at a time $t_{f,i}^{(X)}$.
Then, the 2nd type of AHP current $I_{AHP,i}^{(X)}(t)$ follows after spiking (i.e., $t \geq t_{f,i}^{(X)}$), :
\begin{equation}
I_{AHP,i}^{(X)}(t) = g_{AHP}^{(X)}(t) ~(v_{i}^{(X)}(t) - V_{AHP}^{(X)})~~~{\rm ~for~} \; t \ge t_{f,i}^{(X)}.
\label{eq:AHP1}
\end{equation}
Here, $V_{AHP}^{(X)}$ is the reversal potential for the AHP current, and the conductance $g_{AHP}^{(X)}(t)$ is given by an exponential-decay
function:
\begin{equation}
g_{AHP}^{(X)}(t) = \bar{g}_{AHP}^{(X)}~  e^{-(t-t_{f,i}^{(X)})/\tau_{AHP}^{(X)}},
\label{eq:AHP2}
\end{equation}
where, $\bar{g}_{AHP}^{(X)}$ and $\tau_{AHP}^{(X)}$ are the maximum conductance and the decay time constant for the AHP current.
With increasing $\tau_{AHP}^{(X)}$, the refractory period becomes longer.

The parameter values of the capacitance $C_X$, the leakage current $I_L^{(X)}(t)$, and the AHP current $I_{AHP}^{(X)}(t)$ are the same as those
in the DG networks for sparsely synchronized rhythm and winner-take-all competition in \cite{WTA,SSR}, and refer to Table 1 in \cite{WTA}; these
parameter values are based on physiological properties of the GC, BC, MC, and HIPP cell \cite{Chavlis,Hilus3}.

\subsection{Synaptic Currents in The DG Spiking Neural Network}
\label{subsec:SC}

In Eq.~(\ref{eq:GE}), we consider the synaptic current $I_{syn,i}^{(X)}(t)$ into the $i$th cell in the $X$ population, consisting of the following 3 types of synaptic currents:
\begin{equation}
I_{syn,i}^{(X)}(t) = I_{{\rm AMPA},i}^{(X,Y)}(t) + I_{{\rm NMDA},i}^{(X,Y)}(t) + I_{{\rm GABA},i}^{(X,Z)}(t).
\label{eq:ISyn1}
\end{equation}
Here, $I_{{\rm AMPA},i}^{(X,Y)}(t)$ and $I_{{\rm NMDA},i}^{(X,Y)}(t)$ are the excitatory AMPA ($\alpha$-amino-3-hydroxy-5-methyl-4-isoxazolepropionic acid) receptor-mediated and NMDA ($N$-methyl-$D$-aspartate) receptor-mediated currents from the pre-synaptic source $Y$ population to the post-synaptic $i$th neuron in the target $X$ population, respectively. On the other hand, $I_{{\rm GABA},i}^{(X,Z)}(t)$ is the inhibitory $\rm GABA_A$ ($\gamma$-aminobutyric acid type A) receptor-mediated current from the pre-synaptic source $Z$ population to the post-synaptic $i$th neuron in the target $X$ population.

As in the case of the AHP current, the $R$ (= AMPA, NMDA, or GABA) receptor-mediated synaptic current $I_{R,i}^{(T,S)}(t)$ from the pre-synaptic source $S$ population to the $i$th post-synaptic cell in the target $T$ population is given by:
\begin{equation}
I_{R,i}^{(T,S)}(t) = g_{R,i}^{(T,S)}(t)~(v_{i}^{(T)}(t) - V_{R}^{(S)}),
\label{eq:ISyn2}
\end{equation}
where $g_{(R,i)}^{(T,S)}(t)$ and $V_R^{(S)}$ are synaptic conductance and synaptic reversal potential
(determined by the type of the pre-synaptic source $S$ population), respectively.

In the case of the $R$ (=AMPA and GABA)-mediated synaptic currents, we obtain the synaptic conductance $g_{R,i}^{(T,S)}(t)$ from:
\begin{equation}
g_{R,i}^{(T,S)}(t) = K_{R}^{(T,S)} \sum_{j=1}^{N_S} w_{ij}^{(T,S)} ~ s_{j}^{(T,S)}(t).
\label{eq:ISyn3}
\end{equation}
Here, $K_{R}^{(T,S)}$ is the synaptic strength per synapse for the $R$-mediated synaptic current
from the $j$th pre-synaptic neuron in the source $S$ population to the $i$th post-synaptic cell in the target $T$ population.
The inter-population synaptic connection from the source $S$ population (with $N_s$ cells) to the target $T$ population is given by the connection weight matrix
$W^{(T,S)}$ ($=\{ w_{ij}^{(T,S)} \}$) where $w_{ij}^{(T,S)}=1$ if the $j$th cell in the source $S$ population is pre-synaptic to the $i$th cell
in the target $T$ population; otherwise $w_{ij}^{(T,S)}=0$. The fraction of open ion channels at time $t$ is also represented by $s^{(T,S)}(t)$.

\begin{table*}
\caption{Parameters for the synaptic currents $I_R^{({\rm GC},S)}(t)$ into the GC. The GCs receive the direct excitatory input from the entorhinal cortex (EC) cells, the inhibitory input from the HIPP cells, the excitatory input from the MCs, and the feedback inhibition from the BCs.
}
\label{tab:Synparm1}
\begin{tabular}{|c|c|c|c|c|c|c|}
\hline
Target Cells ($T$) & \multicolumn{6}{c|}{GC} \\
\hline
Source Cells ($S$) & \multicolumn{2}{c|}{EC cell} & HIPP cell & \multicolumn{2}{c|}{MC} & BC \\
\hline
Receptor ($R$) & AMPA & NMDA & GABA & AMPA & NMDA & GABA \\
\hline
$K_{R}^{(T,S)}$ & 0.89 & 0.15 & 0.13 & 0.05 & 0.01 & 25.0 \\
\hline
$\tau_{R,r}^{(T,S)}$ & 0.1 & 0.33 & 0.9 & 0.1 & 0.33 & 0.9 \\
\hline
$\tau_{R,d}^{(T,S)}$ & 2.5 & 50.0 & 6.8 & 2.5 & 50.0 & 6.8 \\
\hline
$\tau_{R,l}^{(T,S)}$ & 3.0 & 3.0 & 1.6 & 3.0 & 3.0 & 0.85 \\
\hline
$V_{R}^{(S)}$ & 0.0 & 0.0 & -86.0 & 0.0 & 0.0 & -86.0 \\
\hline
\end{tabular}
\end{table*}

\begin{table*}
\caption{Parameters for the synaptic currents $I_R^{(T,S)}(t)$ into the HIPP cell, MC, and BC. The HIPP cells receive the excitatory input from the EC cells, the MCs receive the excitatory input from the GCs, and the BCs receive the excitatory inputs from both the GCs and the MCs.
}
\label{tab:Synparm2}
\begin{tabular}{|c|c|c|c|c|c|c|c|c|c|}
\hline
Target Cells ($T$) & \multicolumn{2}{c|}{HIPP cell} & \multicolumn{2}{c|}{MC} & \multicolumn{5}{c|}{BC}\\
\hline
Source Cells ($S$) & \multicolumn{2}{c|}{EC cell} & \multicolumn{2}{c|}{GC} & \multicolumn{2}{c|}{GC} & \multicolumn{2}{c|}{MC} & HIPP cell \\
\hline
Receptor ($R$) & AMPA & NMDA & AMPA & NMDA & AMPA & NMDA & AMPA & NMDA & GABA \\
\hline
$K_{R}^{(T,S)}$ & 12.0 & 3.04 & 7.25 & 1.31 & 1.24 & 0.06 & 5.3 & 0.29 & 8.05 \\
\hline
$\tau_{R,r}^{(T,S)}$ & 2.0 & 4.8 & 0.5 & 4.0 & 2.5 & 10.0 & 2.5 & 10.0 & 0.4 \\
\hline
$\tau_{R,d}^{(T,S)}$ & 11.0 & 110.0 & 6.2 & 100.0 & 3.5 & 130.0 & 3.5 & 130.0 & 5.8 \\
\hline
$\tau_{R,l}^{(T,S)}$ & 3.0 & 3.0 & 1.5 & 1.5 & 0.8 & 0.8 & 3.0 & 3.0 & 1.6 \\
\hline
$V_{R}^{(S)}$ & 0.0 & 0.0 & 0.0 & 0.0 & 0.0 & 0.0 & 0.0 & 0.0 & -86.0 \\
\hline
\end{tabular}
\end{table*}

On the other hand, in the NMDA-receptor case, some of the post-synaptic NMDA channels are blocked by the positive magnesium ion ${\rm Mg}^{2+}$
\cite{NMDA}. Therefore, the conductance in the case of NMDA receptor is given by \cite{Chavlis}:
\begin{equation}
g_{R,i}^{(T,S)}(t) = {\widetilde K}_R^{(T,S)}~f(v^{(T)}(t))~\sum_{j=1}^{N_S} w_{ij}^{(T,S)} ~ s_{j}^{(T,S)}(t).
\label{eq:NMDA}
\end{equation}
Here, ${\widetilde K}_R^{(T,S)}$ is the synaptic strength per synapse, and the fraction of NMDA channels that are not blocked by the ${\rm Mg}^{2+}$ ion is given by a sigmoidal function $f(v^{(T)}(t))$:
\begin{equation}
f(v^{(T)}(t)) = \frac{1}{1+\eta\cdot [{\rm Mg}^{2+}]_o \cdot \exp(-\gamma \cdot v^{(T)}(t))}.
\end{equation}
Here, $v^{(T)}(t)$ is the membrane of the target cell, $[{\rm Mg}^{2+} ]_o$ is the outer ${\rm Mg}^{2+}$ concentration, $\eta$ denotes the sensitivity of ${\rm Mg}^{2+}$ unblock, $\gamma$ represents the steepness of ${\rm Mg}^{2+}$ unblock, and the values of parameters change depending on the target cell \cite{Chavlis}.
For simplicity, some approximation to replace $f(v^{(T)}(t))$ with $\langle f(v^{(T)}(t))\rangle$ [i.e., time-averaged value of $f(v^{(T)}(t))$ in the range of $v^{(T)}(t)$ of the target cell] has been made in \cite{SSR}. Then, an effective synaptic strength $K_{\rm NMDA}^{(T,S)} (={\widetilde K}_{\rm NMDA}^{(T,S)} \langle f(v^{(T)}(t))\rangle$) was introduced by absorbing $\langle f(v^{(T)}(t))\rangle$ into $K_{\rm NMDA}^{(T,S)}$. Thus, with the scaled-down effective synaptic strength $K_{\rm NMDA}^{(T,S)}$ (containing the blockage effect of the ${\rm Mg}^{2+}$ ion), the conductance $g$ for the NMDA receptor may also be well approximated in the same form of conductance as other AMPA and GABA receptors in Eq.~(\ref{eq:ISyn3}). In this way, we obtain all the effective synaptic strengths $K_{\rm NMDA}^{(T,S)}$ from the synaptic strengths $\widetilde{K}_{\rm NMDA}^{(T,S)}$ in \cite{Chavlis} by considering the average blockage effect of the ${\rm Mg}^{2+}$ ion. As a result, we can use the same form of synaptic conductance of Eq.~(\ref{eq:ISyn3}) in all the cases of $R=$ AMPA, NMDA, and GABA.

The post-synaptic ion channels are opened because of binding of neurotransmitters (emitted from the source $S$ population) to receptors in the target
$T$ population. The fraction of open ion channels at time $t$ is denoted by $s^{(T,S)}(t)$. The time course of $s_j^{(T,S)}(t)$ of the $j$th cell
in the source $S$ population is given by a sum of double exponential functions $E_{R}^{(T,S)} (t - t_{f}^{(j)}-\tau_{R,l}^{(T,S)})$:
\begin{equation}
s_{j}^{(T,S)}(t) = \sum_{f=1}^{F_{j}^{(s)}} E_{R}^{(T,S)} (t - t_{f}^{(j)}-\tau_{R,l}^{(T,S)}).
\label{eq:ISyn4}
\end{equation}
Here, $t_f^{(j)}$ and $F_j^{(s)}$ are the $f$th spike time and the total number of spikes of the $j$th cell in the source $S$ population, respectively, and
$\tau_{R,l}^{(T,S)}$ is the synaptic latency time constant for $R$-mediated synaptic current.
The exponential-decay function $E_{R}^{(T,S)} (t)$ (corresponding to contribution of a pre-synaptic spike occurring at $t=0$ in the absence of synaptic latency)
is given by:
\begin{equation}
E_{R}^{(T,S)}(t) = \frac{1}{\tau_{R,d}^{(T,S)}-\tau_{R,r}^{(T,S)}} \left( e^{-t/\tau_{R,d}^{(T,S)}} - e^{-t/\tau_{R,r}^{(T,S)}} \right) \cdot \Theta(t). \label{eq:ISyn5}
\end{equation}
Here, $\Theta(t)$ is the Heaviside step function: $\Theta(t)=1$ for $t \geq 0$ and 0 for $t <0$, and $\tau_{R,r}^{(T,S)}$ and $\tau_{R,d}^{(T,S)}$ are synaptic rising and decay time constants of the $R$-mediated synaptic current, respectively.

In comparison to those in the case of sparsely synchronized rhythms \cite{SSR}, most of the parameter values, related to the synaptic currents, are the same, except for changes in the strengths for the synapses, HIPP $\rightarrow$ GC, GC $\rightarrow$ MC, and MC $\rightarrow$ BC; these changes are made to keep the same activation degree of the GCs as in the case of sparsely synchronized rhythm \cite{SSR}. In the present DG network for the pattern separation,
a new disynaptic connection from HIPP cells to GCs, mediated by BCs, is added, in addition to the (pre-existing) disynaptic path from MCs to GCs in \cite{SSR}.
The strength for the synapse, HIPP $\rightarrow$ BC, is determined, based on the information in \cite{BN1,BN2}.
For completeness, we include Tables \ref{tab:Synparm1} and \ref{tab:Synparm2} which show the parameter values for the synaptic strength per synapse $K_{R}^{(T,S)}$, the synaptic rising time constant $\tau_{R,r}^{(T,S)}$, synaptic decay time constant $\tau_{R,d}^{(T,S)}$, synaptic latency time constant $\tau_{R,l}^{(T,S)}$, and the synaptic reversal potential  $V_{R}^{(S)}$ for the synaptic currents into the GCs and for the synaptic currents into the HIPP cells, the MCs and the BCs, respectively. These parameter values are also based on the physiological properties of the relevant cells \cite{Chavlis,SynParm1,SynParm2,SynParm3,SynParm4,SynParm5,SynParm6,SynParm7,SynParm8}.

All of our source codes for computational works were written in C language. Then, using the GCC compiler we run the source codes on personal computers with CPU (i5-10210U; 1.6 GHz) and 8 GB RAM; the number of used personal computers change (from 1 to 70) depending on the type of jobs. Numerical integration of the governing Eq.~(\ref{eq:GE}) for the time-evolution of states of individual spiking neurons is done by employing the 2nd-order Runge-Kutta method with the time step 0.1 msec. We will release our source codes at the public database such as ModelDB.

\section{Disynaptic Effect of The Hilar Cells on Pattern Separation}
\label{sec:PS}
In this section, we study the disynaptic effect of the excitatory MCs and the inhibitory HIPP cells on pattern separation (performed by the GCs) in our spiking neural network of the hippocampal DG. Disynaptic inhibition from the MCs decreases the firing activity of the GCs, while due to their disinhibition of the BCs, the disynaptic effect of the HIPP cells results in increase in the spiking activity of the GCs. In this way, their disynaptic effects on the GCs are opposite.
As a result of balance between the two competing disynaptic effects of the MCs and the HIPP cells, when simultaneously varying both $K^{\rm (BC, MC)}$ and $K^{\rm (BC, HIPP)}$ from their default values, the pattern separation degree ${\cal S}_d$ is found to form a bell-shaped curve with an optimal maximum at their default values. During the pattern separation, sparsely synchronized rhythm also appears in the population of the GCs.
The amplitude measure ${\cal M}_a$ (representing population synchronization degree) and the random-phase-locking degree ${\cal L}_d$ (denoting individual
firing activity degree) of the sparsely synchronized rhythm of the GCs are found to be correlated with the pattern separation degree ${\cal S}_d$. Hence, the larger the population synchronization degree of the sparsely synchronized rhythm becomes, the more the pattern separation becomes enhanced.

\subsection{Characterization of Pattern Separation by Varying The Overlap Percentage between The Two Input Patterns}
\label{subsec:POL}
The EC (entorhinal cortex) is the external source providing the excitatory inputs to the GCs and the HIPP cells via the PPs (perforant paths) \cite{Myers1,Myers2,Myers3,Scharfman,Chavlis,WTA,SSR}. The activation degree $D_a^{(in)}$ of the input EC cells is chosen as 10 $\%$ \cite{ANA4}.
Thus, we randomly choose 40 active ones among the 400 EC cells. Each active EC cell is modeled in terms of the Poisson spike train with frequency of 40 Hz \cite{ANA5}. We characterize pattern separation between the input and the output patterns via integration of the
governing equations (\ref{eq:GE}). In each realization, we have a break stage (0-300 msec) (for which the network reaches a stable state), and then a stimulus stage (300-1,300 msec) follows; the stimulus period $T_s$ (for which network analysis is done) is 1,000 msec. During the stimulus stage, we obtain the output firings of the GCs. For characterization of pattern separation between the input and the output patterns, 30 realizations are done.

The input (spiking) patterns of the 400 EC cells and the output (spiking) patterns of the 2,000 GCs are given in terms of binary representations \cite{Myers1,Chavlis}; active and silent cells are represented by 1 and 0, respectively. Here, active cells show at least one spike during the stimulus stage; otherwise, silent cells. In each realization, we first make a random choice of an input pattern $A^{(in)}$ for the EC cells, and then construct another input patterns $B_i^{(in)}$ ($i=1,\dots,9$) from the base input pattern $A^{(in)}$ with the overlap percentage $P_{OL}$ $=90, \dots,$ and $10$ $\%$, respectively, in the following way \cite{Myers1,Chavlis}. Among the active EC cells in the pattern $A^{(in)}$, we randomly choose active cells for the pattern $B_i^{(in)}$ with the probability $P_{OL}~\%$ (e.g., in the case of $P_{OL} = 80 ~\%$, we randomly choose 32 active EC cells among the 40 active EC cells in the base pattern $A^{(in)}$).
The remaining active EC cells in the pattern $B_i^{(in)}$ are randomly chosen in the subgroup of silent EC cells in the pattern $A^{(in)}$
(e.g., for  $P_{OL} = 80 ~\%$, 8 additional active EC cells in the pattern $B_i^{(in)}$ are randomly chosen in the subgroup of 360 silent EC cells in the pattern
$A^{(in)}$).

\begin{figure}
\includegraphics[width=0.9\columnwidth]{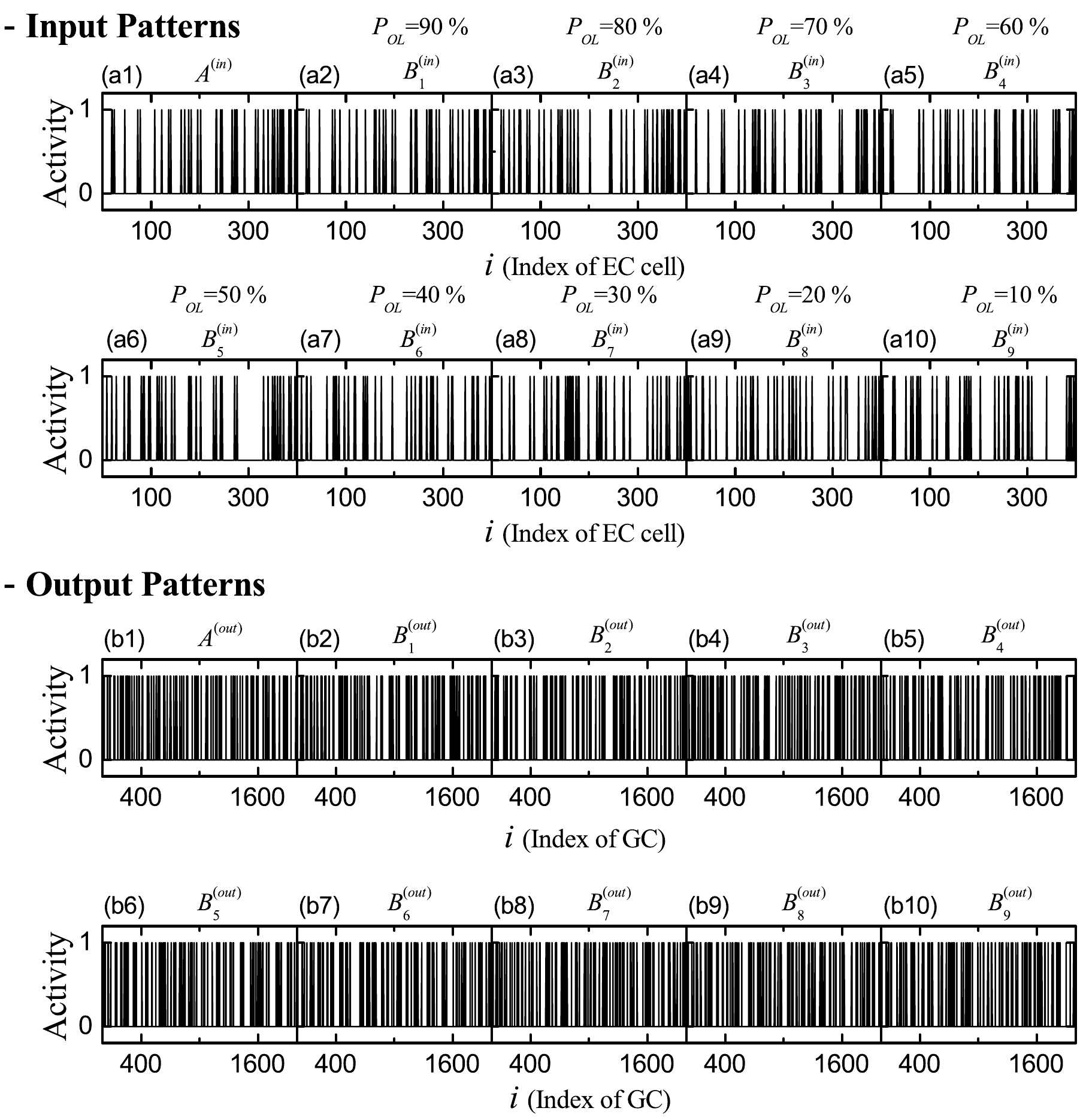}
\caption{Binary-representation plots of spiking activity for (a1)-(a10) the input and (b1)-(b10) the output patterns for 9 values of overlap percentage $P_{OL}$.
}
\label{fig:SAP}
\end{figure}

Figure \ref{fig:SAP}(a1)-\ref{fig:SAP}(a10) show binary-representation plots of spiking activity of the 400 EC cells [active (silent) cell: 1 (0)] for the input patterns $A^{(in)}$ and $B_i^{(in)}$ ($i=1, \dots , 9)$ in the case of 9 values of overlap percentage $P_{OL}$.
Through integration of the governing equations (\ref{eq:GE}), we also get the output patterns $A^{(out)}$ and $B_i^{(out)}$ ($i=1, \dots, 9)$ of the 2,000 GCs
for the input patterns $A^{(in)}$ and $B_i^{(in)}$ ($i=1, \dots, 9),$ respectively. The binary-representation plots of spiking activity of the 2,000 GCs
for the output patterns $A^{(out)}$ and $B_i^{(out)}$ ($i=1, \dots, 9)$  are shown in Figs.~\ref{fig:SAP}(b1)-\ref{fig:SAP}(b10), respectively.

From now on, we characterize pattern separation between the input and the output patterns by changing the overlap percentage $P_{OL}$.
For a pair of patterns, $A^{(x)}$ and $B^{(x)}$, the pattern distance $D_p^{(x)}$ between the two input ($x=in)$ or output ($x=out$) patterns is given by \cite{Chavlis}:
\begin{equation}
D_p^{(x)} = {\frac {O^{(x)}} {D_a^{(x)}} }.
\label{eq:PD}
\end{equation}
Here, $D_a^{(x)}$ is the average activation degree of the two patterns $A^{(x)}$ and $B^{(x)}$:
\begin{equation}
D_a^{(x)} = {\frac {(D_a^{(A^{(x)})} + D_a^{(B^{(x)})})} {2} },
\label{eq:AD}
\end{equation}
and $O^{(x)}$ is the orthogonalization degree between the patterns $A^{(x)}$ and $B^{(x)}$, representing their ``dissimilarity'' degree.
As the average activation degree is lower (i.e., more sparse firing) and the orthogonalization degree is higher (i.e., more dissimilar), their pattern distance increases.

\begin{figure}
\includegraphics[width=0.9\columnwidth]{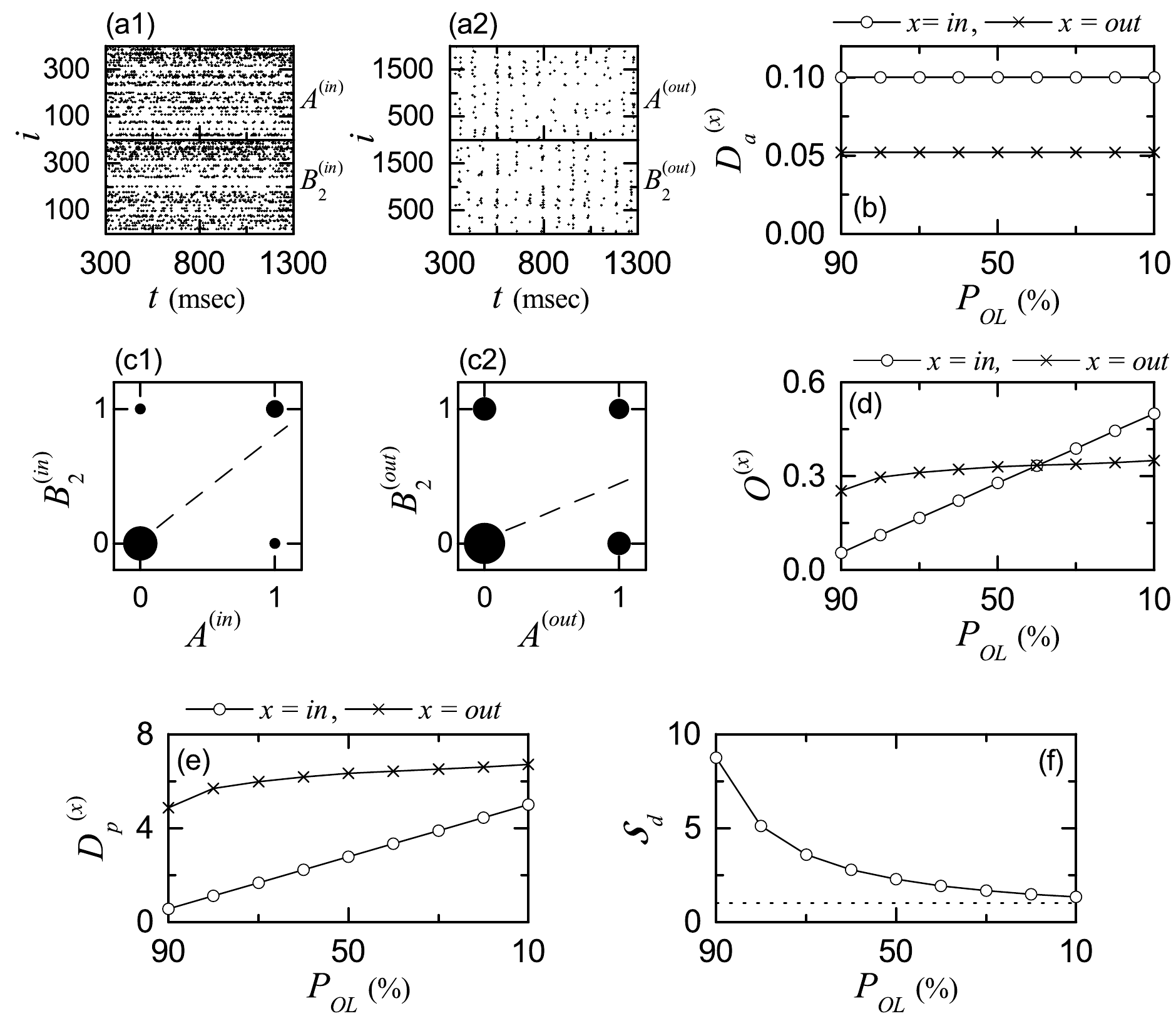}
\caption{Characterization of pattern separation between the input and the output patterns.
(a1) Raster plots of spikes of ECs for the input patterns $A^{(in)}$ and $B^{(in)}_2$ in the case of overlap percentage $P_{OL}=80 \%$.
(a2) Raster plots of spikes of GCs for the output patterns $A^{(out)}$ and $B^{(out)}_2$.
(b) Plots of average activation degree $D_a^{(x)}$ versus $P_{OL}$ for the input ($x=in$; open circle) and the output ($x=out$, cross) patterns.
Plots of the diagonal elements (0, 0) and (1, 1) and the anti-diagonal elements (1, 0) and (0, 1) for the spiking activity (1: active; 0: silent)
in the pair of (c1) input ($x=in$) and (c2) output ($x=out$) patterns $A^{(x)}$ and $B^{(x)}_2$ for $P_{OL}=80\%$;
sizes of solid circles, located at (0,0), (1,1), (1,0), and (0,1), are given by the integer obtained by rounding off the number of $5~ \log_{10} (n_p)$ ($n_p$: number of data at each location), and a dashed linear least-squares fitted line is also given. (d) Plots of average orthogonalization degree $O^{(x)}$ versus $P_{OL}$ in the case of the input ($x=in$; open circle) and the output ($x=out$, cross) patterns. (e) Plots of the pattern distance $D_p^{(x)}$ versus $P_{OL}$ for the input ($x=in$; open circle) and the output ($x=out$, cross) patterns. (g) Plots of pattern separation degrees ${\cal S}_d$ versus $P_{OL}$.
}
\label{fig:POL}
\end{figure}

Let $\{ a^{(x)}_i \}$ and $\{ b^{(x)}_i \}$ ($i=1,\dots,N_x$) be the binary representations [1 (0) for the active (silent) cell] of the two input ($x=in$) or output ($x=out$) spiking patterns $A^{(x)}$ and $B^{(x)}$, respectively; $N_{in}=N_{\rm EC}=400$ and $N_{out}=N_{\rm GC}=2,000$.
Then, the Pearson's correlation coefficient $\rho^{(x)},$ denoting the ``similarity'' degree between the two patterns, is given by
\begin{equation}
\rho^{(x)} = \frac{\sum_{i=1}^{N_x}\Delta a^{(x)}_i \cdot \Delta b^{(x)}_i}{\sqrt{\sum_{i=1}^{N_x}\Delta {a^{(x)}_i}^2}\sqrt{\sum_{i=1}^{N_x}\Delta {b^{(x)}_i}^2}}.
\label{eq:Rho}
\end{equation}
Here, $\Delta a^{(x)}_i = a^{(x)}_i - \langle {a^{(x)}_i} \rangle$, $\Delta b^{(x)}_i = b^{(x)}_i - \langle {b^{(x)}_i} \rangle$, and
$\langle \cdots \rangle$ denotes population average over all cells; the range of $\rho^{(x)}$ is [-1, 1].  Then, the orthogonalization degree $O^{(x)},$ representing the dissimilarity degree between the two patterns, is given by:
\begin{equation}
O^{(x)} = {\frac {(1-\rho^{(x)})} {2} },
\label{eq:OD}
\end{equation}
where the range of $O^{(x)}$ is [0, 1].
With $D_a^{(x)}$ and $O^{(x)}$, we may get the pattern distances of Eq.~(\ref{eq:PD}), $D_p^{(in)}$ and $D_p^{(out)}$, for the input and the output pattern pairs, respectively. Then, the pattern separation degree ${\cal S}_d $ is given by the ratio of $D_p^{(out)}$ to $D_p^{(in)}$:
\begin{equation}
{\cal S}_d = {\frac {D_p^{(out)}} {D_p^{(in)}} }.
\label{eq:PSD}
\end{equation}
If ${\cal S}_d > 1$, the output pattern pair of the GCs is more dissimilar than the input pattern pair of the EC cells, which results in
occurrence of pattern separation.

Figure \ref{fig:POL}(a1) shows the raster plots of spikes of 400 EC cells (i.e. a collection of spike trains of individual EC cells) for the input patterns
$A^{(in)}$ and $B_2^{(in)}$ in the case of $P_{OL}=80~\%$. In this case, the activation degree $D_a^{(in)}$ is chosen as 10 $\%$, independently of the input patterns. Figure \ref{fig:POL}(a2) shows the raster plots of spikes of 2,000 GCs for the output patterns $A^{(out)}$ and $B_2^{(out)}$. As shown well in the raster plots of spikes, the GCs exhibit more sparse firings than the EC cells. In this case, the average activation degree of Eq.~(\ref{eq:AD}), $D_a^{(out)}$, is 5.2 $\%$
(which is obtained via 30 realizations). Figure \ref{fig:POL}(b) shows the plot of the average activation degree $D_a^{(x)}$ versus the overlap percentage $P_{OL}$;
open circles denote the case of input patterns ($x=in$) and crosses represent the case of output patterns ($x=out$). We note that $D_a^{(out)} = 0.052$ (i.e., 5.2
$\%$), independently of $P_{OL}$. Then, the sparsity ratio, ${\cal R}_s$ ($= D_a^{(in)} / D_a^{(out)}$), becomes 1.923; the output patterns are 1.923 times as sparse
as the input patterns.

Figures \ref{fig:POL}(c1) and (c2) show plots of the diagonal elements (0, 0) and (1, 1) and the anti-diagonal elements (1, 0) and (0, 1) for the spiking activity (1: active; 0: silent) in the pair of input ($x=in$) and output ($x=out$) patterns $A^{(x)}$ and $B^{(x)}_2$ for $P_{OL}=80\%$, respectively.
In each plot, the sizes of solid circles, located at (0,0), (1,1), (1,0), and (0,1), are given by the integer obtained by rounding off the number of  $5~ \log_{10} (n_p)$ ($n_p$: number of data at each location), and a dashed linear least-squares fitted line is also given. In this case, the Pearson's correlation coefficients of Eq.~(\ref{eq:Rho}) (obtained via 30 realizations) for the pairs of the input and the output patterns are $\rho^{(in)}=0.7778$ and $\rho^{(out)}=0.4118$, which correspond to the slopes of the dashed fitted lines.
Then, from Eq.~(\ref{eq:OD}), we get the average orthogonalization degrees for the pairs of the input and the output patterns: $O^{(in)}=0.1111$ and
$O^{(out)}=0.2941$.

Figure \ref{fig:POL}(d) shows plots of average orthogonalization degree $O^{(x)}$ versus $P_{OL}$ in the case of the input ($x=in$; open circle) and the output ($x=out$, cross) patterns. In the case of the pairs of the input patterns, with decreasing $P_{OL}$ from 90 $\%$ to 10 $\%$, $O^{(in)}$ increases linearly
from 0.0556 to 0.5. In the case of the pairs of the output patterns, $O^{(out)}$ begins from a larger value (0.2536), but slowly increases to 0.3498 for
$P_{OL}=10$ $\%$ (which is lower than $O^{(in)}$). Thus, the two lines of $O^{(in)}$ and $O^{(out)}$ cross for $P_{OL} \simeq 40$ $\%$.
Hence, for $P_{OL}$ less than 40 $\%$, $O^{(out)}$ is larger than $O^{(in)}$ (i.e., the pair of output patterns is more dissimilar than the pair of input patterns).
In contrast, for $P_{OL} > 40~\%$, the pair of output patterns becomes less dissimilar than the pair of input patterns, because $O^{(in)}$  is larger than
$O^{(out)}$.

With the average activation degrees $D_a^{(x)}$ and the average orthogonalization degrees $O^{(x)}$, we can get the pattern distances $D_p^{(x)}$
of Eq.~(\ref{eq:PD}) for the pairs of input and the output patterns. Figure \ref{fig:POL}(e) shows plots of the pattern distance $D_p^{(x)}$ versus $P_{OL}$
in the case of the input ($x=in$; open circle) and the output ($x=out$, cross) patterns. We note that, for all values of $P_{OL}$,
$D_p^{(out)} >  D_p^{(in)}$ (i.e., the pattern distance for the pair of output patterns is larger than that for the pair of input patterns).
However, the difference between $D_p^{(out)}$ and $D_p^{(in)}$ is found to decrease, as the overlap percentage $P_{OL}$ is decreased.
(i.e., the pair of the input patterns has larger pattern distance $D_p^{(in)}$).

Finally, we get the pattern separation degree ${\cal S}_d$ of Eq.~(\ref{eq:PSD}) through the ratio of $D_p^{(out)}$ to $D_p^{(in)}$.
Figure \ref{fig:POL}(f) shows plots of the pattern separation degree ${\cal S}_d$ versus $P_{OL}$. As $P_{OL}$ is decreased from
90 $\%$ to 0, ${\cal S}_d$ is found to decrease from 8.7623 to 1.3432. Hence, for all values of $P_{OL}$, pattern separation occurs because ${\cal S}_d >1$.
However, the smaller $P_{OL}$ is, the lower ${\cal S}_d$ becomes.

\begin{figure}
\includegraphics[width=1.0\columnwidth]{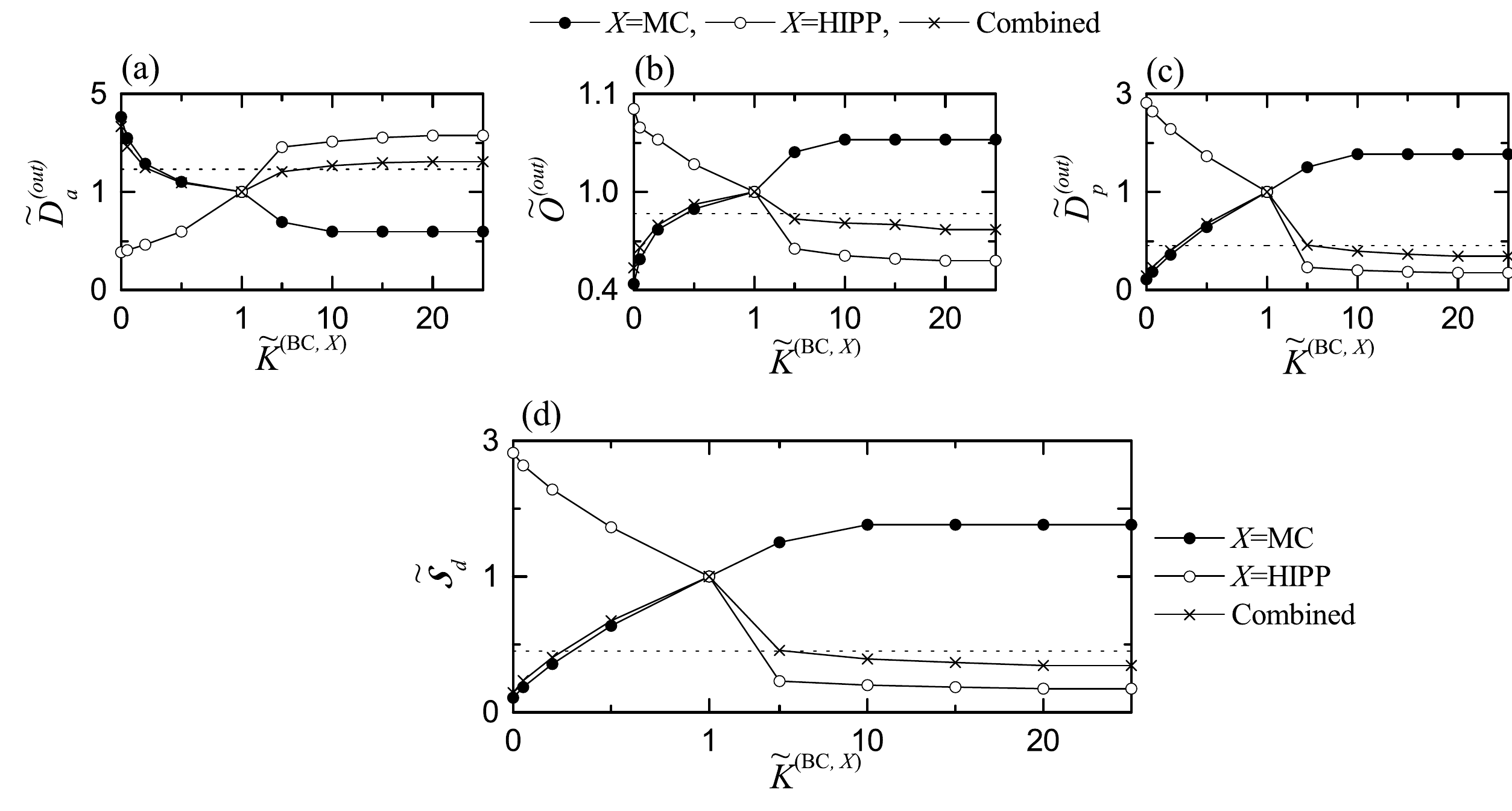}
\caption{Disynaptic effect of the MCs and the HIPP cells on pattern separation. Plots of (a) the normalized average activation degree ${\widetilde D}_a^{(out)},$ (b) the normalized average orthogonalization degree ${\widetilde O}^{(out)},$ and (c) the normalized pattern distance ${\widetilde D}_p^{(out)}$ versus the normalized synaptic strength ${\widetilde K}^{\rm {(BC,X)}}$ (X= MC or HIPP) for the output patterns. (d) Plot of the normalized pattern separation degree ${\widetilde {\cal S}}_d$ versus ${\widetilde K}^{\rm {(BC,X)}}$. In (a)-(d), solid circles, open circles, and crosses represent the cases of the MCs and the HIPP cells and the combined case, respectively. For clear presentation, we choose four different scales around (1, 1); (left, right) and (up, down).
The horizontal dotted lines in (a)-(c) represent ${\widetilde D}_a^{(in)}$ (normalized average activation degree), ${\widetilde O}^{(in)}$ (normalized orthogonalization degree), and ${\widetilde D}_p^{(in)}$ (normalized pattern distance) for the input patterns, respectively.
The horizontal dotted lines in (d) denotes a threshold value of ${\widetilde {\cal S}}_d^* \simeq 0.451$ (corresponding to ${\cal S}_d=1$).
}
\label{fig:PS}
\end{figure}

\subsection{Disynaptic Effect of The MCs and The HIPP Cells on Pattern Separation}
\label{subsec:DSE}
Here, we use the normalized synaptic strength ${\widetilde K}_{R}^{\rm (BC,X)}$ (= $K_{R}^{\rm (BC,X)} / {K_{R}^{\rm (BC,X)}}^*$) ($\rm X=$ MC or HIPP; $R=$ AMPA, NMDA, or GABA). ${K_{R}^{\rm (BC,X)}}^*$ is the original default value; ${K_{\rm AMPA}^{\rm (BC,MC)}}^* = 5.3$, ${K_{\rm NMDA}^{\rm (BC,MC)}}^* = 0.29,$ and
${K_{\rm GABA}^{\rm (BC,HIPP)}}^* = 8.05$. We change ${\widetilde K}_{\rm AMPA}^{\rm (BC,MC)}$ and ${\widetilde K}_{\rm NMDA}^{\rm (BC,MC)}$ in the same way such that ${\widetilde K}_{\rm AMPA}^{\rm (BC,MC)} = {\widetilde K}_{\rm NMDA}^{\rm (BC,MC)} \equiv {\widetilde K}^{\rm (BC,MC)}$, and investigate the disynaptic effect of the MCs on pattern separation. Similarly, we vary ${\widetilde K}_{\rm GABA}^{\rm (BC,HIPP)}$ (for brevity, we write it as
${\widetilde K}^{\rm (BC,HIPP)}$), and investigate the disynaptic effect of the HIPP cells on pattern separation.

In each realization for a given ${\widetilde K}^{\rm (BC,X)}$ (X= MC or HIPP), we consider 9 pairs of input patterns $(A^{(in)}, B_i^{(in)})$ $(i=1,\dots, 9$) with the overlap percentage $P_{OL}= 90,\dots,$ and $10~ \%$, respectively. All quantities for the input patterns are independent of ${\widetilde K}^{\rm (BC,X)}$. The activation degree $D_a^{(in)}$ is 0.1 (10 $\%$), independently of the pairs. For each pair $(A^{(in)}, B_i^{(in)})$, we get the realization-averaged orthogonalization degree $\langle O^{(in)}(i) \rangle_r$ (denoting the dissimilarity degree between the patterns) via 30 realizations. Then, the average orthogonalization degree $O^{(in)}~(=0.2778)$ for the input-pattern pairs is given by the average of $\langle O^{(in)}(i) \rangle_r$ over all pairs. In this way, we get $O^{(in)}$ via double averaging processes (i.e., realization and pair averaging). Then, the pattern distance $D_p^{(in)}$ of Eq.~(\ref{eq:PD}) between the two input patterns (given by the ratio of the average orthogonalization degree to the average activation degree) becomes 2.778. The normalized quantities ${\widetilde D}_a^{(in)}$, ${\widetilde O}^{(in)}$, and ${\widetilde D}_p^{(in)}$ [divided by ${D_a^{(out)}}^*$, ${O^{(out)}}^*$, and ${D_p^{(out)}}^*$
(average values for the output-pattern pairs at the default values ${K_{R}^{\rm (BC,X)}}^*$)] are represented by the horizontal dotted lines in Figs.~\ref{fig:PS}(a)-\ref{fig:PS}(c), respectively.

As in the case of the input-pattern pairs, we get $D_a^{(out)}$, $O^{(out)}$, and $D_p^{(out)}$ of the output-pattern pairs for each ${\widetilde K}^{\rm (BC,X)}$ (X= MC or HIPP) through double averaging processes (i.e., realization and pair averaging).
We first study the disynaptic effect of the MCs on the activation degree $D_a^{(out)}$ of the output patterns by varying ${\widetilde K}^{\rm (BC,MC)}$.
With decreasing ${\widetilde K}^{\rm (BC,MC)}$ from 1 (i.e., default values), the average activation degree $D_a^{(out)}$ of Eq.~(\ref{eq:AD})
is found to increase from 5.2 $\%$ to 22.1 $\%$. As ${\widetilde K}^{\rm (BC,MC)}$  is increased from 1, $D_a^{(out)}$ is found to decrease from 5.2 $\%$ and
becomes saturated to 3.1 $\%$ at ${\widetilde K}^{\rm (BC,MC)} \sim 10$. Here, we introduce normalized average activation degree ${\widetilde D}_a^{(out)}$
[= $D_a^{(out)} / {D_a^{(out)}}^*$]; ${D_a^{(out)}}^*$ (= 5.2 $\%$) is the average activation degree at the default values, ${K_{R}^{\rm (BC,MC)}}^*$.
Then, as ${K_{R}^{\rm (BC,MC)}}^*$ is increased from 0, the disynaptic effect of the MCs (reducing the firing activity of the GCs) increases, and hence
${\widetilde D}_a^{(out)}$ is found to decrease from 4.058 and become saturated to 0.596 for ${K_{R}^{\rm (BC,MC)}}^* \sim 10$, as shown in Fig.~\ref{fig:PS}(a)
(solid circles).

Next, we study the disynaptic effect of the HIPP cells (disinhibiting the BCs) on $D_a^{(out)}$ by changing ${\widetilde K}^{\rm (BC,HIPP)}$.
As shown in Fig.~\ref{fig:PS}(a) (open circles), with increasing ${\widetilde K}^{\rm (BC,HIPP)}$ from 0, the disynaptic effect of the HIPP cells (enhancing the firing activity of the GCs via increased disinhibition of the BCs) increases, and hence ${\widetilde D}_a^{(out)}$ is found to increase from 0.385 and become saturated to 3.288 for ${K_{R}^{\rm (BC,MC)}}^* \sim 20$.

We note that the disynaptic effects of the MCs and the HIPP cells are opposite ones. Then, we consider a combined case where we simultaneously change both
${\widetilde K}^{\rm (BC,MC)}$ and ${\widetilde K}^{\rm (BC,HIPP)}$ such that ${\widetilde K}^{\rm (BC,MC)} = {\widetilde K}^{\rm (BC,HIPP)} \equiv
{\widetilde K}^{\rm (BC,X)}$. As a result of balance between the two competing disynaptic effects of the MCs and the HIPP cells, ${\widetilde D}_a^{(out)}$ is found to form a well-shaped curve (crosses) with an optimal minimum at ${\widetilde K}^{\rm (BC,X)} = 1$ (i.e., at the default values), as shown in Fig.~\ref{fig:PS}(a).

In addition to the average activation degree $D_a^{(out)}$, we consider the disynaptic effects of the MCs and the HIPP cells on the average orthogonalization degree $O^{(out)}$ (representing the dissimilarity degree between the output patterns) and the pattern distance $D_p^{(out)}$ (given by the ratio of the average orthogonalization degree to the average activation degree) for the output patterns. At the original default values ${K_{R}^{\rm (BC,X)}}^*$, the average orthogonalization degree ${O^{(out)}}^*$ is 0.320 and the pattern distance ${D_p^{(out)}}^*$ (given by $O^{(out)} / D_a^{(out)}$) is 6.154. By changing the normalized synaptic strength ${\widetilde K}^{\rm (BC,X)}$ (X= MC or HIPP), we study the disynaptic effects of the MCs and the HIPP cells on $O^{(out)}$  and $D_p^{(out)}$, which are well shown in Figs.~\ref{fig:PS}(b) and \ref{fig:PS}(c), respectively; solid circles (variation in ${\widetilde K}^{\rm (BC,MC)}$), open circles (change in ${\widetilde K}^{\rm (BC,HIPP)}$), and crosses (combined case with change in both ${\widetilde K}^{\rm (BC,MC)}$ and ${\widetilde K}^{\rm (BC,HIPP)}$). As in the case of $D_a^{(out)}$, we use the normalized average orthogonalization degree ${\widetilde O}^{(out)}$ ($= O^{(out)} / {O^{(out)}}^*$) and the normalized pattern distance ${\widetilde {D_p}}^{(out)}$  ($= {D_p}^{(out)} / {{D_p}^{(out)}}^*$).

With increasing ${\widetilde K}^{\rm (BC,MC)}$ from 0, ${\widetilde O}^{(out)}$  is found to increase from 0.438 and get saturated to 1.053 at ${\widetilde K}^{\rm (BC,MC)} \sim 10$, in contrast to the case of ${\widetilde {D}}_a^{(out)}$. On the other hand, as ${\widetilde K}^{\rm (BC,HIPP)}$ is increased from 0,
${\widetilde O}^{(out)}$ is found to decrease from 1.084 and get saturated to 0.578 at ${\widetilde K}^{\rm (BC,HIPP)} \sim 20$.
In the combined case where both ${\widetilde K}^{\rm (BC,MC)}$ and  ${\widetilde K}^{\rm (BC,HIPP)}$ are simultaneously changed,
${\widetilde O}^{(out)}$ is found to form a bell-shaped curve at an optimal maximum at ${\widetilde K}^{\rm (BC,X)} = 1$ (i.e., at the default values),
in contrast to the case of ${\widetilde {D}}_a^{(out)}$ with a well-shaped curve.

We note that the normalized pattern distance ${\widetilde {D}}_p^{(out)}$ in Fig.~\ref{fig:PS}(c) is found to exhibit the same kind of changing tendency as ${\widetilde O}^{(out)}$ in Fig.~\ref{fig:PS}(b). The disynaptic effects of the MCs and the HIPP cells are opposite ones; with increasing
${\widetilde K}^{\rm (BC,MC)}$ (${\widetilde K}^{\rm (BC,HIPP)})$, ${\widetilde {D}}_p^{(out)}$ is found to increase (decrease).
As a result of balance between the competing disynaptic effects of the MCs and the HIPP cells, in the combined case of simultaneous change in both
${\widetilde K}^{\rm (BC,MC)}$ and ${\widetilde K}^{\rm (BC,HIPP)}$, ${\widetilde {D}}_p^{(out)}$ is found to form a bell-shaped curve with an optimal
maximum at ${\widetilde K}^{\rm (BC,X)} = 1$.

Finally, we investigate the disynaptic effect of the MCs and the HIPP cells on the pattern separation degree ${\cal S}_d$ of Eq.~(\ref{eq:PSD})
(given by the ratio of $D_p^{(out)}$ to $D_p^{(in)}$). At the original default values ${K_{R}^{\rm (BC,X)}}^*$, ${\cal S}_d^*$ is 2.215.
As ${\widetilde K}^{\rm (BC,MC)}$ (${\widetilde K}^{\rm (BC,HIPP)}$) is increased, the normalized pattern separation degree
$\widetilde{ {\cal S}}_d$ is found to increase (decrease) from 0.108 (2.819) and to become saturated to 1.767 (0.176) at
${\widetilde K}^{\rm (BC,MC)} \sim 10$ (${\widetilde K}^{\rm (BC,HIPP)} \sim 20$).
In the combined case where both ${\widetilde K}^{\rm (BC,MC)}$ and ${\widetilde K}^{\rm (BC,HIPP)}$ are simultaneously changed,
it is found that, $\widetilde{ {\cal S}}_d$ forms a bell-shaped curve with an optimal maximum for ${\widetilde K}^{\rm (BC,X)} = 1$.
The horizontal dotted line in Fig.~\ref{fig:PS}(d) represents a threshold value of ${\widetilde {\cal S}}_d^* \simeq 0.451$ (corresponding to
${\cal S}_d=1$). We note that pattern separation may occur only when ${\cal S}_d > 1$ (i.e., ${\widetilde {\cal S}}_d > {\widetilde {\cal S}}_d^*$);
otherwise, no pattern separation occurs because $D_p^{(in)} >  D_p^{(out)}$.
Hence, in the combined case, pattern separation occurs for $0.25 < {\widetilde K}^{\rm (BC,X)} < 5.4$.
For ${\widetilde K}^{\rm (BC,X)} < 0.25$, pattern separation cannot occur because the disynaptic effect of the MCs is so much decreased.
On the other hand, for ${\widetilde K}^{\rm (BC,X)} > 5.4$, no pattern separation occurs due to so much increased disynaptic effect of the HIPP cells.

\subsection{Quantitative Association between Sparsely Synchronized Rhythm of the GCs and Pattern Separation}
\label{subsec:SSR}
While the GCs perform pattern separation, sparsely synchronized rhythm is found to appear in the population of the GCs \cite{SSR}.
In the combined case of simultaneous change in both ${\widetilde K}^{\rm (BC,MC)}$ and ${\widetilde K}^{\rm (BC,HIPP)}$, we investigate how the population
and the individual behaviors in the sparsely synchronized rhythm are changed. Here, we consider a long-term stimulus stage (300-30,300 msec) (i.e., the stimulus period $T_s=30,000$ msec) without realization, in contrast to the case of pattern separation with short-term stimulus period (1,000 msec) and 30 realizations, because long-term stimulus is necessary for analysis of dynamical behaviors.

\begin{figure}
\includegraphics[width=1.0\columnwidth]{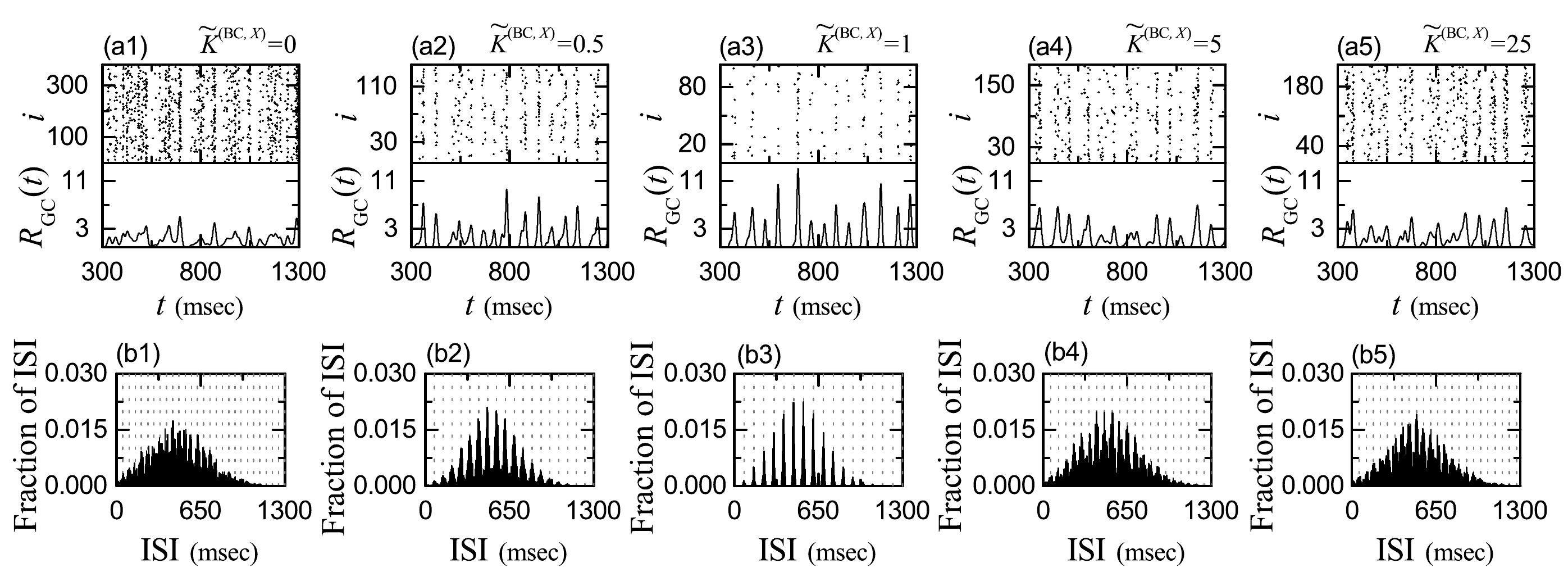}
\caption{Sparsely synchronized rhythms of the active GCs and Multi-peaked ISI histograms. (a1)-(a5) Raster plots of spikes and IPSRs $R_{\rm GC}(t)$ for the active GCs for ${\widetilde K}^{\rm (BC,X)}$ (X= MC or HIPP) = 0, 0.5, 1.0, 5.0 and 25, respectively.
(b1)-(b5) Population-averaged ISI histograms for ${\widetilde K}^{\rm (BC,X)}$ (X= MC or HIPP) = 0, 0.5, 1.0, 5.0 and 25, respectively;
bin size = 2 msec. Vertical dotted lines in (b1)-(b5) represent the integer multiples of the global period $T_G^{\rm (GC)}$ of  $R_{\rm GC}(t)$.
}
\label{fig:SSR1}
\end{figure}

Population firing activity of the active GCs may be well visualized in the raster plot of spikes which is a collection of spike trains of individual active GCs.
Figures \ref{fig:SSR1}(a1)-\ref{fig:SSR1}(a5) show the raster plots of spikes for the active GCs for ${\widetilde K}^{\rm (BC,X)}$ (X= MC or HIPP) = 0, 0.5, 1.0, 5.0 and 25, respectively. For convenience, only a part from $t=300$ to 1,300 msec is shown in each raster plot of spikes. We note that sparsely synchronized stripes (composed of sparse spikes and indicating population sparse synchronization) appear successively.

As a population quantity showing collective behaviors, we employ an IPSR (instantaneous population spike rate) which may be obtained from the raster plot of spikes
\cite{W_Review,Sparse1,Sparse2,Sparse3,FSS,SM}. To get a smooth IPSR, we employ the kernel density estimation (kernel smoother) \cite{Kernel}. Each spike in the raster plot is convoluted (or blurred) with a kernel function $K_h(t)$ to get a smooth estimate of IPSR $R_{\rm GC}(t)$:
\begin{equation}
R_{\rm{GC}}(t) = \frac{1}{N_a} \sum_{i=1}^{N_a} \sum_{s=1}^{n_i} K_h (t-t_{s}^{(i)}),
\label{eq:IPSR}
\end{equation}
where $N_a$ is the number of the active GCs, $t_{s}^{(i)}$ is the $s$th spiking time of the $i$th active GC, $n_i$ is the total number of spikes for the $i$th active GC, and we use a Gaussian kernel function of band width $h$:
\begin{equation}
K_h (t) = \frac{1}{\sqrt{2\pi}h} e^{-t^2 / 2h^2}, ~~~~ -\infty < t < \infty,
\label{eq:Gaussian}
\end{equation}
where the band width $h$ of $K_h(t)$ is 20 msec. The IPSRs $R_{\rm {GC}}(t)$ of the active GCs are also shown in Figs.~\ref{fig:SSR1}(a1)-\ref{fig:SSR1}(a5) for ${\widetilde K}^{\rm (BC,X)}$ (X= MC or HIPP) = 0, 0.5, 1.0, 5.0 and 25, respectively.

We note that the IPSRs $R_{\rm {GC}}(t)$ exhibit synchronous oscillations.
However, as ${\widetilde K}^{\rm (BC,X)}$ is changed (i.e., increased or decreased) from 1 (i.e., original default value), the amplitude of $R_{\rm {GC}}(t)$, representing the synchronization degree of the sparsely synchronized rhythm, makes a distinct decrease, mainly because the pacing degree between spikes in each spiking stripe in the rater plot of spikes becomes worse. In this way, the synchronization degree of the sparsely synchronized rhythm becomes maximal at the default value ${\widetilde K}^{\rm (BC,X)}=1$ (i.e., as ${\widetilde K}^{\rm (BC,X)}$ is changed from 1, the synchronization degree of the sparsely synchronized rhythm
is decreased).

In addition to the population firing behavior, we also consider the individual spiking behavior of the active GCs. We obtain the inter-spike-interval (ISI) histogram for each active GC by collecting the ISIs during the stimulus period $T_s$ ($= 3 \cdot 10^4$ msec), and then get the population-averaged ISI histogram by averaging the individual ISI histograms for all the active GCs. Figures \ref{fig:SSR1}(b1)-\ref{fig:SSR1}(b5) show the population-averaged ISI histograms for ${\widetilde K}^{\rm (BC,X)}$ (X= MC or HIPP) = 0, 0.5, 1.0, 5.0 and 15, respectively.

Each active GC exhibits intermittent spikings, phase-locked to $R_{\rm GC}(t)$ at random multiples of its global period $T_G^{\rm (GC)}$. Due to the random spike skipping, distinct multiple peaks appear at the integer multiples of $T_G^{\rm (GC)}$ (denoted by the vertical dotted lines) in the ISI histogram. This is in contrast to the case of full synchronization where only one dominant peak appears at the global period $T_G^{\rm (GC)}$; all cells fire regularly at each global cycle without skipping. Hereafter, these peaks will be called as the random-spike-skipping peaks.

\begin{figure}
\includegraphics[width=0.8\columnwidth]{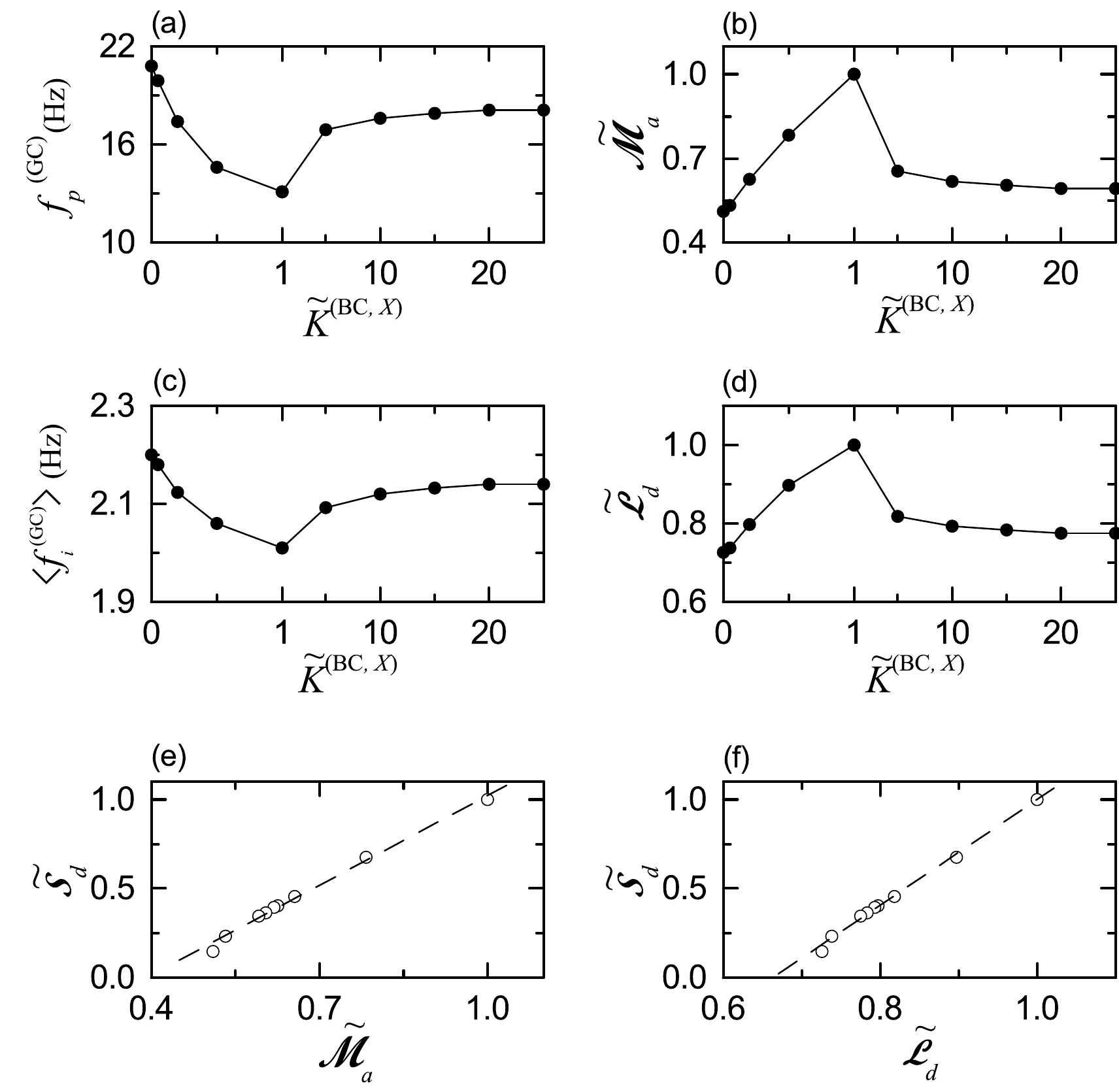}
\caption{Quantitative relationship between sparsely synchronized rhythm of the GCs and pattern separation in the combined case of
simultaneously changing the normalized synaptic strengths ${\widetilde K}^{\rm {(BC,MC)}}$ and ${\widetilde K}^{\rm {(BC,HIPP)}}$.
(a) Plot of the population frequency $f_p^{(GC)}$ of sparsely synchronized rhythm of the GCs
    versus ${\widetilde K}^{\rm {(BC,X)}}$ (X= MC or HIPP).
(b) Plot of the normalized amplitude measure, ${\widetilde {\cal M}}_a$ versus ${\widetilde K}^{\rm {(BC,X)}}$.
(c) Plot of the population-averaged mean firing rate $\langle f_i \rangle$ versus ${\widetilde K}^{\rm {(BC,X)}}$.
(d) Plot of the normalized random-spike-skipping degree ${\widetilde {\cal L}}_d$ versus ${\widetilde K}^{\rm {(BC,X)}}$.
(e) Plot of the normalized pattern separation degree ${\widetilde {\cal S}}_d$ versus ${\widetilde {\cal M}}_a$.
(f) Plot of the normalized pattern separation degree ${\widetilde {\cal S}}_d$ versus ${\widetilde {\cal L}}_d$.
Dashed fitted lines are given in (e)-(f).
}
\label{fig:SSR2}
\end{figure}

In the default case of ${\widetilde K}^{\rm (BC,X)}=1,$ there appear 13 distinct clear random-spike-skipping peaks in the ISI histogram of Fig.~\ref{fig:SSR1}(c3).
The middle 6th- and 7th-order peaks are the highest ones, and hence spiking may occur most probably after 5- or 6-times spike skipping. This kind of structure in the ISI histogram is a little different from that in the case of fast sparse synchronization where the highest peak appears at the 1st-order peak, and then the heights of the higher-order peaks decrease successively \cite{W_Review,Sparse1,Sparse2,Sparse3,FSS}.

As ${\widetilde K}^{\rm (BC,X)}$ is changed (i.e., increased or decreased) from 1, the random-spike-skipping peaks become smeared more and more, along with decrease
in the height of the highest peak and appearance of higher-order peaks. Thus, with increasing or decreasing ${\widetilde K}^{\rm (BC,X)}$,
the random phase-locking degree, representing how well intermittent spikes make phase-locking to $R_{\rm GC}(t)$ at random multiples of its global period $T_G^{\rm (GC)}$, is decreased.

From now on, in Figs.~\ref{fig:SSR2}(a)-\ref{fig:SSR2}(d), we make quantitative characterization of population and individual firing behaviors in the sparsely synchronized rhythm of the GCs. Figures \ref{fig:SSR2}(a) and \ref{fig:SSR2}(c) show the plots of the population frequency $f_p^{(\rm GC)}$ [i.e., the oscillating frequency of $R_{\rm GC}(t)$] of the sparsely synchronized rhythm and the population-averaged mean firing rate (MFR) $\langle f_i^{(\rm GC)} \rangle$
of individual active GCs, respectively. At the default value of ${\widetilde K}^{\rm (BC,X)} = 1$, the population-averaged MFR $\langle f_i^{\rm (GC)} \rangle$ (= 2.01 Hz) is much less than the population frequency $f_p^{\rm (GC)}~(=13.1$ Hz) for the sparsely synchronized rhythm, due to random spike skipping, which is in contrast to the case of full synchronization where the population-averaged MFR is the same as the population frequency.

As ${\widetilde K}^{\rm (BC,X)}$ is decreased from 1, the disynaptic inhibition effect of the MCs becomes dominant and
decreased (i.e., less exciting the BCs). Hence, the firing activity of the GCs is found to increase, which results in increase in both $f_p^{\rm (GC)}$ and $\langle f_i^{\rm (GC)} \rangle$. On the other hand, with increasing ${\widetilde K}^{\rm (BC,X)}$ from 1, the disynaptic effect of the HIPP cells becomes dominant and
increased (i.e, more disinhibiting the BCs). Therefore, the firing activity of the GCs is also found to increase, which leads to increase in both $f_p^{\rm (GC)}$ and $\langle f_i^{\rm (GC)} \rangle$.

We characterize the synchronization degree of the sparsely synchronized rhythm of the GCs in terms of the amplitude measure ${\cal M}_a$, given by the time-averaged amplitude of the IPSR $R_{\rm {GC}}(t)$ \cite{AM}:
\begin{equation}
  {\cal M}_a = {\overline {{\cal A}_i}};~ {\cal A}_i = \frac { [R_{\rm GC, max}^{(i)}(t) - R_{\rm GC, min}^{(i)}(t)] } {2},
\label{eq:AM}
\end{equation}
where the overline represents time average, and $R_{\rm GC,max}^{(i)}(t)$ and $R_{\rm GC,min}^{(i)}(t)$ are the maximum and the minimum of $R_{\rm GC}(t)$ in its $i$th global cycle (corresponding to the $i$th spiking stripe), respectively. As ${\cal M}_a$ increases (i.e., the time-averaged amplitude of $R_{\rm GC}(t)$ is increased), the synchronization degree of the sparsely synchronized becomes higher.

Figure \ref{fig:SSR2}(b) shows the plot of the normalized amplitude measure ${\widetilde {\cal M}}_a$ (= ${\cal M}_a / {\cal M}_a^*$) versus
${\widetilde K}^{\rm (BC,X)}$; ${\cal M}_a^*$ (= 3.566) is the default value for ${\widetilde K}^{\rm (BC,X)} = 1$.
We note that the normalized amplitude measure ${\widetilde {\cal M}}_a$ is found to form a bell-shaped curve with an optimal maximum at the default value
${\widetilde K}^{\rm (BC,X)} = 1$. As ${\widetilde K}^{\rm (BC,X)}$ is decreased from 1, ${\widetilde {\cal M}}_a$ is decreased from 1 to 0.510, because
the disynaptic inhibition effect of the MCs becomes dominant and decreased. Also, with increasing ${\widetilde K}^{\rm (BC,X)}$ from 1, ${\widetilde {\cal M}}_a$ is also decreased from 1 and becomes saturated to 0.592 for ${\widetilde K}^{\rm (BC,X)} \sim 20$, because the disynaptic effect of the HIPP cells becomes dominant and increased.

Next, we characterize the individual spiking behavior of the active GCs in the ISI histogram with multiple peaks resulting from random spike skipping.
We introduced a new random phase-locking degree ${\cal L}_d$, denoting how well intermittent spikes make phase-locking to $R_{\rm GC}(t)$ at random multiples of its global period $T_G^{\rm (GC)}$ \cite{SSR}, and characterize the degree of random spike skipping seen in the ISI histogram in terms of ${\cal L}_d$. By following the approach developed in the case of pacing degree between spikes in the stripes in the raster plot of spikes \cite{SM}, the random phase-locking degree was introduced to examine the regularity of individual firings (represented well in the sharpness of the random-spike-skipping peaks).

We first locate the random-spike-skipping peaks in the ISI histogram. For each $n$th-order peak, we get the normalized weight $w_n$, given by:
\begin{equation}
w_n = \frac {N_{\rm ISI}^{(n)}} {N_{\rm ISI}^{(tot)}},
\label{eq:Wn}
\end{equation}
where $N_{\rm ISI}^{(tot)}$ is the total number of ISIs obtained during the stimulus period ($T_s$ $= 3 \cdot 10^4$ msec) and $N_{\rm ISI}^{(n)}$ is the number of the ISIs in the $n$th-order peak.

We now consider the sequence of the ISIs, $\{ {\rm ISI}_i^{(n)},~ i=1,\dots,N_{\rm ISI}^{(n)} \}$, within the $n$th-order peak, and get the random
phase-locking degree ${\cal L}_d^{(n)}$ of the $n$th-order peak. Similar to the case of the pacing degree between spikes \cite{SM}, we provide a phase $\psi$ to each ${\rm ISI}_i^{(n)}$ via linear interpolation; for details, refer to \cite{SSR}. Then, the contribution of the ${\rm ISI}_i^{(n)}$ to the locking degree ${\cal L}_d^{(n)}$ is given by $\cos ( \psi_i^{(n)})$; $\psi_i^{(n)}$ denotes the phase for ${\rm ISI}_i^{(n)}$.
An ${\rm ISI}_i^{(n)}$ makes the most constructive contribution to ${\cal L}_d^{(n)}$ for $\psi_i^{(n)}=0$, while it makes no contribution to ${\cal L}_d^{(n)}$ for $\psi = {\frac {\pi} {2}}$ or $-{\frac {\pi} {2}}$. By averaging the matching contributions of all the ISIs in the $n$th-order peak, we get:
\begin{equation}
  {\cal L}_d^{(n)} = { \frac {1}{N_{\rm ISI}^{(n)}} } \sum_i^{N_{\rm ISI}^{(n)}}  \cos ( \psi_i^{(n)}).
\label{eq:LDn}
\end{equation}

Then, we obtain the (overall) random phase-locking degree ${\cal L}_d$ via weighted average of the random phase-locking degrees ${\cal L}_d^{(n)}$ of all the peaks:
\begin{equation}
  {\cal L}_d = \sum_{n=1}^{N_p} w_n \cdot {\cal L}_d^{(n)}
             = {\frac {1} {N_{\rm ISI}^{\rm (tot)}} } \sum_{n=1}^{N_p} \sum_{i=1}^{N_{\rm ISI}^{\rm (tot)} } \cos ( \psi_i^{(n)}),
\label{eq:LD}
\end{equation}
where $N_p$ is the number of peaks in the ISI histogram.
Thus, ${\cal L}_d$ corresponds to the average of contributions of all the ISIs in the ISI histogram.

Figure \ref{fig:SSR2}(d) shows the plot of the normalized random phase-locking degree ${\widetilde {\cal L}}_d$ (=${\cal L}_d / {\cal L}_d^*$)  versus
${\widetilde K}^{\rm (BC,X)}$; ${\cal L}_d^*$ (= 0.911) is the default value for ${\widetilde K}^{\rm (BC,X)} = 1$.
We note that the normalized random phase-locking degree ${\widetilde {\cal L}}_d$ is found to form a bell-shaped curve with an optimal maximum at the default value
of ${\widetilde K}^{\rm (BC,X)} = 1$. In the default case, the random phase-locking degree ${\cal L}_d^*~(=0.911)$, characterizing the sharpness of all the peaks, is  high. Hence, the GCs make intermittent spikes which are well phase-locked to $R_{\rm GC}(t)$ at random multiples of its global period $T_G^{\rm (GC)}$.
However, with decreasing ${\widetilde K}^{\rm (BC,X)}$ from 1, ${\widetilde {\cal L}}_d$ is decreased from 1 to 0.726, because the decreased disynaptic inhibition effect of the MCs becomes dominant. Also, as ${\widetilde K}^{\rm (BC,X)}$ is increased from 1, ${\widetilde {\cal M}}_a$ is decreased from 1 and becomes saturated to 0.775 for ${\widetilde K}^{\rm (BC,X)} \sim 20$, because the increased disynaptic effect of the HIPP cells becomes dominant.

Finally, we investigate quantitative association between sparsely synchronized rhythm and pattern separation.
Figures \ref{fig:SSR2}(e) and \ref{fig:SSR2}(f) show plots of ${\widetilde {\cal M}}_a$ and ${\widetilde {\cal L}}_d$ versus the normalized pattern separation degree  ${\widetilde {\cal S}}_d$, respectively. Population (${\widetilde {\cal M}}_a$) and individual (${\widetilde {\cal L}}_d$) firing behaviors in the sparsely synchronized rhythm of the GCs (performing pattern separation) are found to be positively correlated with the pattern separation (${\widetilde {\cal S}}_d$) with the Pearson's correlation coefficients $r=0.9959$ and 0.9975, respectively. Hence, as the synchronization and the random phase-locking degrees in the sparsely synchronized rhythm of the GCs become larger, the pattern separation degree becomes higher.

\section{Summary and Discussion}
\label{sec:SUM}
We considered the disynaptic paths from the hilar cells (i.e., excitatory MCs and inhibitory HIPP cells) to the principal excitatory GCs (performing pattern separation), mediated by the inhibitory BCs in our spiking neural network of the hippocampal DG; MC $\rightarrow$ BC $\rightarrow$ GC and HIPP $\rightarrow$ BC $\rightarrow$ GC. We note that, disynaptic inhibition from the MCs decreases the activity of the GCs, while due to their disinhibition of the BCs, the disynaptic effect of the HIPP cells leads to increase in the activity of the GCs. In this way, their disynaptic effects on the GCs are opposite.

By changing the synaptic strength $K^{\rm (BC, X)}$ [from the presynaptic X (= MC or HIPP) to the postsynaptic BC] from the default value ${K^{\rm (BC, X)}}^*$, we investigated the disynaptic effect of the MCs and the HIPP cells on the pattern separation (transforming the input patterns from the EC into sparser and orthogonalized output patterns) performed by the principal GCs. The pattern-separated output patterns are projected to the pyramidal cells in the CA3 subregion, which facilitates pattern storage and retrieval in the CA3. In this way, the DG plays a role of pre-processor for the CA3.

We first studied the disynaptic effect of the MCs by varying the normalized synaptic strength $\widetilde{K}^{\rm (BC, MC)}$ [= $K^{\rm (BC, MC)}~/~{K^{\rm (BC, MC)}}^*$]; ${K^{\rm (BC, MC)}}^*$ is the original default value. When $\widetilde{K}^{\rm (BC, MC)}$ was decreased from 1 (i.e., default value) to 0, the normalized pattern separation degree ${\widetilde {\cal S}}_d$ [= ${\cal S}_d~/~{\cal S}_d^*$] (${\cal S}_d^*$: pattern separation degree at the default value) was decreased from 1 to 0.108. Such decrease in ${\widetilde {\cal S}}_d$ was found to result from increase in the activation degree $D_a^{(out)}$ of the GCs and decrease in the orthogonalization degree $O^{(out)}$ between the two output patterns (generated by the GCs). In contrast, as $\widetilde{K}^{\rm (BC, MC)}$ was increased from 1, ${\widetilde {\cal S}}_d$ began to increase from 1 and become saturated to ${\widetilde {\cal S}}_d \simeq 1.767$ for $\widetilde{K}^{\rm (BC, MC)} \sim 10$. In this case, decrease in the activation degree $D_a^{(out)}$ of the GCs and increase in the orthogonalization degree $O^{(out)}$ was found to lead to increase in ${\widetilde {\cal S}}_d$. Overall, in the whole range of $\widetilde{K}^{\rm (BC, MC)}$, as it is increased from 0, the normalized pattern separation degree ${\widetilde {\cal S}}_d$ was found to increase from 0.108 and get saturated to 1.767 for $\widetilde{K}^{\rm (BC, MC)} \sim 10$ [see Fig.~\ref{fig:PS}(d)].

Next, we studied the disynaptic effect of the HIPP cells by varying the normalized synaptic strength $\widetilde{K}^{\rm (BC, HIPP)}$.
The HIPP cells disinhibit the BCs, in contrast to the case of the MCs enhancing the activity of the BCs.
Hence, the disynaptic effect of the HIPP cells was found to be opposite to that of the MCs.
Overall, as it is increased from 0, the normalized pattern separation degree ${\widetilde {\cal S}}_d$ was found to decrease from 2.819 and get saturated
to 0.176 for $\widetilde{K}^{\rm (BC, HIPP)} \sim 20$ [see Fig.~\ref{fig:PS}(d)], because of increase in the activation degree $D_a^{(out)}$ of the GCs and decrease in the orthogonalization degree $O^{(out)}$.

As a 3rd step, we considered the combined case when the two normalized synaptic strengths $\widetilde{K}^{\rm (BC, MC)}$ and $\widetilde{K}^{\rm (BC, HIPP)}$ are changed simultaneously. As they were increased from 0, the normalized pattern separation degree ${\widetilde {\cal S}}_d$ was found to form a bell-shaped curve with an optimal maximum at the default values (i.e., $\widetilde{K}^{\rm (BC, MC)} = \widetilde{K}^{\rm (BC, HIPP)} = 1$). In this combined case, the normalized activation degree ${\widetilde{D}}_a^{(out)}$ was found to form a well-shaped curve with an optimal minimum at the default values, while the normalized orthogonalization degree ${\widetilde O}^{(out)}$ was found to form a bell-shaped curve with an optimal maximum at the default values. In this way, as a result of balance between the competing disynaptic effects of the MCs and the HIPP cells, the pattern separation degree became optimally maximal at the default values.

Moreover, we also investigated the population and the individual behaviors in the sparsely synchronized rhythm of the GCs.
The amplitude measure ${\cal M}_a$ (representing population synchronization degree) and the random-phase-locking degree ${\cal L}_d$ (denoting individual activity degree) of the sparsely synchronized rhythm were found to be strongly correlated with the pattern separation degree ${\cal S}_d$. Hence, the larger the synchronization and the random-phase-locking degrees in the sparsely synchronized rhythm become, the more the pattern separation becomes enhanced.

\begin{figure}
\includegraphics[width=1.0\columnwidth]{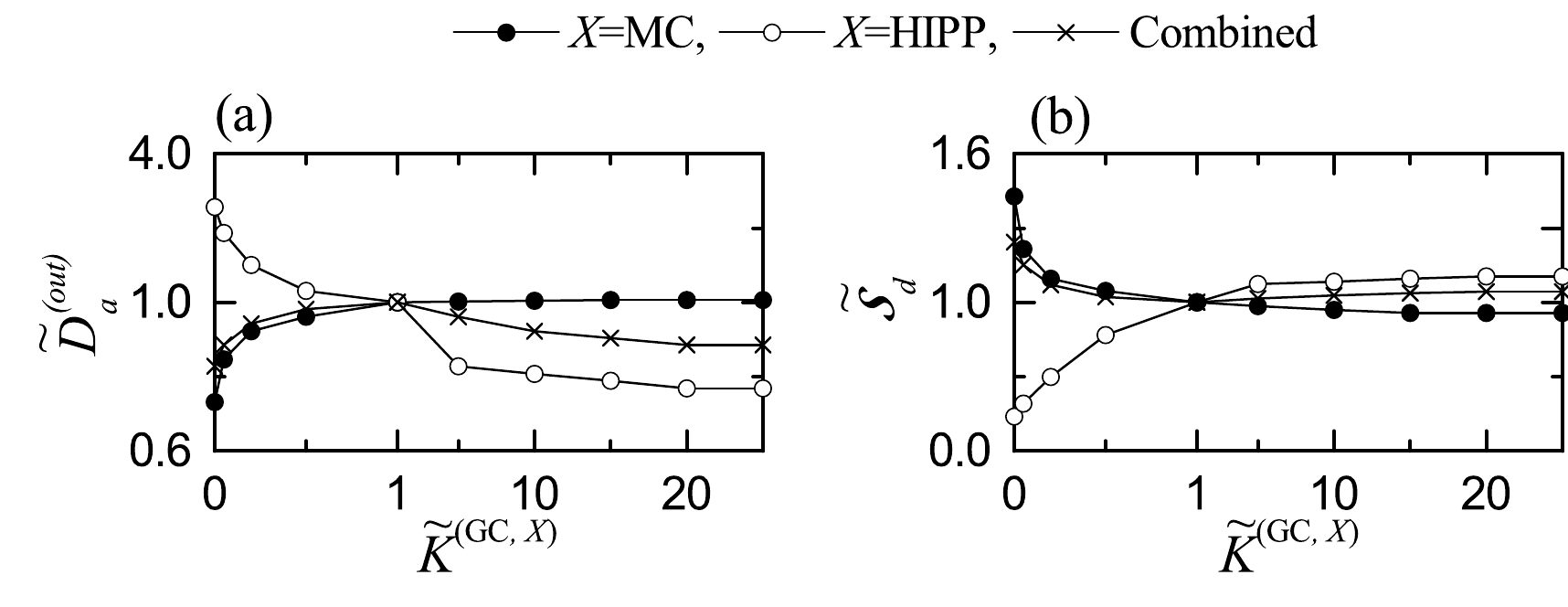}
\caption{Monosynaptic effect of the MCs and the HIPP cells on pattern separation. Plots of (a) the normalized average activation degree ${\widetilde D}_a^{(out)}$ and (b) the normalized pattern separation degree ${\widetilde {\cal S}}_d$ versus the normalized synaptic strength ${\widetilde K}^{\rm {(GC,X)}}$ (X= MC or HIPP) for the output patterns. In (a) and (b), solid circles, open circles, and crosses represent the cases of the MCs and the HIPP cells and the combined case, respectively. For clear presentation, we choose four different scales around (1, 1); (left, right) and (up, down).
}
\label{fig:Mono}
\end{figure}

For comparison, we studied the monosynaptic effect of the MCs and the HIPP cells; MC $\rightarrow$ GC and HIPP $\rightarrow$ GC.
Unlike the disynaptic case, the MCs and the HIPP cells provide direct excitation and inhibition to the GCs, respectively.
Figures \ref{fig:Mono}(a) and \ref{fig:Mono}(b) show the plots of the normalized activation degree ${\widetilde{D}}_a^{(out)}$ of the GCs and the normalized
pattern separation degree ${\widetilde {\cal S}}_d$ versus the normalized synaptic strength $\widetilde{K}^{\rm (GC, X)}$ (X= MC or HIPP), respectively.
Here, the normalized synaptic strength ${\widetilde K}_{R}^{\rm (GC,X)}$ is given by $K_{R}^{\rm (GC,X)} / {K_{R}^{\rm (GC,X)}}^*$ ($\rm X=$ MC or HIPP and $R=$ AMPA, NMDA, or GABA), and ${K_{R}^{\rm (GC,X)}}^*$ is the original default value; ${K_{\rm AMPA}^{\rm (GC,MC)}}^* = 0.05$, ${K_{\rm NMDA}^{\rm (GC,MC)}}^* = 0.01,$ and ${K_{\rm GABA}^{\rm (GC,HIPP)}}^* = 0.13$. In the case of the MCs, we change ${\widetilde K}_{\rm AMPA}^{\rm (GC,MC)}$ and ${\widetilde K}_{\rm NMDA}^{\rm (GC,MC)}$ in the same way such that ${\widetilde K}_{\rm AMPA}^{\rm (GC,MC)} = {\widetilde K}_{\rm NMDA}^{\rm (GC,MC)} \equiv {\widetilde K}^{\rm (GC,MC)}$, and in the case of the HIPP cells, for brevity, we write ${\widetilde K}_{\rm GABA}^{\rm (GC,HIPP)}$ as ${\widetilde K}^{\rm (GC,HIPP)}$.
In the combined case (see crosses in Fig.~\ref{fig:Mono}) for simultaneous change in both $\widetilde{K}^{\rm (GC, MC)}$ and $\widetilde{K}^{\rm (GC, HIPP)}$,
${\widetilde {\cal S}}_d$ (${\widetilde{D}}_a^{(out)}$) was found to form a well-shaped (bell-shaped) curve with an optimal minimum (maximum), in contrast to disynaptic case with the up-down flipped curves. In this combined case, the monosynaptic effect is opposite to the disynaptic effect
(i.e., in the monosynaptic case. the pattern separation degree becomes the lowest at the default value with the highest activation degree).

Finally, we discuss limitations of our present work and future works.
In the present work, although positive correlation between the pattern separation degree and the population synchronization and the random-phase-locking degrees in the sparsely synchronized rhythm of the GCs was found, this kind of correlation does not imply causal relationship. Hence, in future work, it would be interesting to make intensive investigation on their dynamical causation. Also, in the present work, we studied disynaptic effect only in the case of changing the synaptic strength $K^{\rm (BC, X)}$ (X= MC or HIPP). However, in future, it would also be interesting to study disynaptic effect by varying the connection probability $p^{\rm (BC, X)}$ from the presynaptic X to the postsynaptic BC. The effect of decrease in $p^{\rm (BC, X)}$ would be similar to that of decreasing $K^{\rm (BC, X)}$, because the synaptic inputs into the BCs are decreased in both cases. Furthermore, we note that the pyramidal cells in the CA3 provide backprojections to the GCs via polysynaptic connections \cite{Myers2,Myers3,Scharfman}. For example, the pyramidal cells send disynaptic inhibition to the GCs, mediated by the BCs and the HIPP cells in the DG, and they provide trisynaptic inputs to the GCs, mediated by the MCs (pyramidal cells $\rightarrow$ MC $\rightarrow$ BC or HIPP $\rightarrow$ GC). These inhibitory backprojections may decrease the activation degree of the GCs, leading to improvement of pattern separation. Hence, in future work, it would be meaningful to take into consideration the backprojection for the study of pattern separation in the combined DG-CA3 network. Also, in the present study, for simplicity, we did not consider the lamellar organization for the hilar MCs and the HIPP cells, as in \cite{Myers1,Chavlis}; here, we considered only the GC lamellar clusters.
For more refined DG network, it would be necessary in future work to take into consideration the lamellar organization for the MCs and the HIPP cells;
particularly, in the combined DG-CA3 network for pattern storage and retrieval, as in \cite{Myers2,Myers3,Scharfman}.

\section*{Acknowledgments}
This research was supported by the Basic Science Research Program through the National Research Foundation of Korea (NRF) funded by the Ministry of Education (Grant No. 20162007688).

\end{document}